\documentclass[a4paper,11pt]{article}

\usepackage{jcappub} 
\usepackage{aasmacros}
                     
\usepackage[export]{adjustbox} 
\usepackage[utf8]{inputenc}
\usepackage{color}
\usepackage{graphicx}
\usepackage{amsmath,amsfonts,amssymb}

\usepackage{xspace}
\usepackage{multirow}
\usepackage{tabularx}
\usepackage{orcidlink}
\usepackage{ragged2e}
\usepackage{enumitem}
\usepackage[most]{tcolorbox}
\tcbset{sharp corners, boxrule=.4pt, colback=white!98!black,colframe=white!50!black}

\usepackage{cleveref}

\usepackage{setspace}

\usepackage{comment}

\usepackage{listings}
\usepackage[ruled,vlined]{algorithm2e}

\SetCommentSty{mycommfont}

\newcommand{\bea}{\begin{equation}\begin{aligned}} 
\newcommand{\eea}{\end{aligned}\end{equation}}
\newcommand{\be}{\begin{equation}}
\newcommand{\ee}{\end{equation}}

\newcommand{\td}{{\rm d}}

\definecolor{rossocorsa}{rgb}{0.83, 0.0, 0.0}
\definecolor{tardisblue}{rgb}{0.0, 0.18, 0.53}

\title{A non-perturbative framework for $N$-point functions of locally non-Gaussian fields}

\author[a]{Hardi~Veerm\"ae}
\affiliation[a]{Keemilise ja Bioloogilise F\"u\"usika Instituut, R\"avala pst. 10, 10143 Tallinn, Estonia}
\emailAdd{hardi.veermae@cern.ch}

\abstract{We present a non-perturbative approach to correlation functions and polyspectra of locally non-Gaussian fields and develop a semi-perturbative framework that does not rely on a local expansion. This enables the computation of $N$-point functions of non-Gaussian fields even when the mapping between the auxiliary Gaussian field $\zeta_{\rm G}$ and the non-Gaussian field $\zeta({\bf x}) = F(\zeta_{\rm G}({\bf x}))$ is non-analytic. As an example, we consider non-Gaussian fields with exponentially tailed distributions, which can arise, for instance, in ultra-slow-roll models of inflation, and derive some exact analytic results in the strongly non-Gaussian regime.}

\begin{document}
\maketitle
\flushbottom

\section{Introduction}
\label{sec:intro}

Cosmic structures have their origins in random fields. The prevalent picture of the early Universe is that of a hot, expanding Universe with tiny curvature perturbations on an otherwise homogeneous and isotropic Friedmann-Robertson-Walker background. These fluctuations are the seeds of present-day astronomical structures; they are visible in the cosmic microwave background (CMB)~\cite{Planck:2018jri,Planck:2018vyg,Planck:2019kim}, and on smaller scales, they can source cosmological GW backgrounds in the form of scalar-induced gravitational waves (SIGWs)~\cite{Tomita:1967wkp,Matarrese:1992rp,Bruni:1996im, Matarrese:1997ay, Ananda:2006af,Baumann:2007zm,  Domenech:2021ztg} or collapse into primordial black holes (PBHs)~\cite{Zeldovich:1967lct,Hawking:1971ei,Carr:1974nx,Carr:1975qj}, thereby participating in the genesis of dark matter~\cite{Carr:2016drx,Carr:2020xqk,Carr:2026hot}. All these phenomena depend on the statistical characteristics of curvature perturbations.

CMB observations show that primordial curvature perturbations were tiny and nearly Gaussian~\cite{Planck:2019kim}. As long as the non-Gaussianity (NG) remains small, many problems, such as the formation of cosmic structures, PBHs, or scalar-induced GWs, are analytically tractable. Weak NG effects can be treated perturbatively, but a fully non-perturbative treatment is currently limited to lattice studies~\cite{Caravano:2024moy,Caravano:2025diq,Choi:2025eqn}, although some approximate non-perturbative estimates exist, indicating that the perturbative approaches may break down earlier than anticipated for SIGWs~\cite{Iovino:2024sgs}. Thus, even if exact estimates of observably interesting quantities are out of reach computationally, having exact statistical quantities at hand can assist in testing the validity of the perturbative regime. $N$-point functions and the corresponding polyspectra are potential candidates for this.

This study aims to expand the non-perturbative toolbox for dealing with non-Gaussian fields. We will focus on cases where the NG is local, that is, when the NG random field $\zeta({\bf x})$ can be constructed from a Gaussian one $\zeta_{\rm G}({\bf x})$ and the relation depends only on the field at a given point in space, that is,  $\zeta({\bf x}) = F(\zeta_{\rm G}({\bf x}))$, where $F$ is some non-linear real function. In this setup, it is natural to study the abstract properties of the non-Gaussian field in coordinate space and then translate it into Fourier space, which is more commonly used when describing the statistical characteristics of $\zeta({\bf x})$, such as its spectra or bispectra. This approach allows us to provide a simple but rigorous exact formulation for $N$-point correlation functions. Starting from the exact formulation, we will construct a series expansion in the powers of the Gaussian correlation function without relying on the expansion of $F(\zeta_{\rm G}({\bf x}))$. As a result, it is possible to obtain a semi-perturbative treatment applicable in cases where $F(\zeta_{\rm G})$ is non-analytic or where a series expansion of $F(\zeta_{\rm G})$ is not viable.

The main motivation of this work is early universe cosmology, where NG may be strong at small scales and affect the scalar-induced GWs and the formation of PBHs. The results are fairly general, as they hold for generic locally non-Gaussian fields. They can serve as non-perturbative approximations in cases where the field admits a local description over a range of scales of interest. 

The paper is structured as follows: In~\cref{sec:exact and abstract}, we provide definitions and present a simple exact formulation of $N$-point functions. In~\cref{sec:perturbations}, a perturbative expansion to all orders is derived, and a method for effectively resumming the series coefficients is provided. As an example, a class of theoretically well-motivated models of locally non-Gaussian fields with an exponential tail is considered, and its effect on the power spectrum is examined in~\cref{sec:examples}. We conclude in~\cref{sec:concl}.

\section{N-point functions of non-Gaussian fields}
\label{sec:exact and abstract}

A Gaussian field $\zeta_{\rm G}({\bf x})$ with a vanishing mean is completely characterised by its two-point function\footnote{We will denote Gaussian fields and variables related to them by the subindex ${G}$.}
\be\label{eq:def:xi}
    \langle \zeta_{\rm G}({\bf x}_1) \zeta_{\rm G}({\bf x}_2)\rangle 
    = \xi({\bf x}_1,{\bf x}_2)\,.
\ee
Defining the operator $\hat \xi_{\rm G}$ by $(\hat \xi_{\rm G} \zeta_{\rm G})({\bf x}) = \int \td^3 y  \, \xi_{\rm G} ({\bf x},{\bf y}) \zeta_{\rm G}({\bf y})$ and its inverse $\hat\xi^{-1}$ through $\hat\xi_{\rm G} \hat\xi_{\rm G}^{-1} = 1$, then any expectation values of a functional $\mathcal{F}[\zeta_{\rm G}]$ of $\zeta_{\rm G}$ can then be expressed via the statistical path integral
\be\label{eq:expectation_path}
    \langle \mathcal{F}[\zeta_{\rm G}] \rangle 
    = \frac{1}{Z} \int D\zeta_{\rm G} \, \mathcal{F}[\zeta_{\rm G}] \exp\left[  - \frac{1}{2} \zeta_{\rm G} \hat\xi_{\rm G}^{-1} \zeta_{\rm G}  \right]\,,
\ee
where the normalisation factor $Z = (\det 2\pi \hat \xi_{\rm G})^{1/2}$ is given by the functional determinant of $\hat \xi_{\rm G}$. As usual, the path integral \eqref{eq:expectation_path} is understood as the continuum limit of multidimensional Gaussian integrals in the discretised spacetime. In that context, $\hat \xi_{\rm G}$ directly generalises the notion of a covariance matrix of a multivariate Gaussian distribution to the field $\zeta_{\rm G}$. We further note that such stochastic path integrals can be estimated numerically and are better behaved than their quantum counterparts due to their exponential damping at large $\zeta_{\rm G}$.

We will consider cases where the non-Gaussian field $\zeta$ can be constructed from a Gaussian field $\zeta_{\rm G}$ via a general non-linear (and possibly non-local) functional $\mathcal{F}$
\be
    \zeta({\bf x}) = \mathcal{F}[\zeta_{\rm G};{\bf x}]\,.
\ee
The $n$-point function of such fields can be estimated using the path integral \eqref{eq:expectation_path}, that is, as
\bea\label{eq:n_point_general}
    \left\langle \prod_i \zeta({\bf x}_i) \right\rangle 
    &= 
    \frac{1}{Z} 
    \int D\zeta_{\rm G}  \, \left[ \prod_i \mathcal{F}[\zeta_{\rm G};{\bf x}_i] \right]\, e^{- \frac{1}{2}\zeta_{\rm G} \hat\xi^{-1} \zeta_{\rm G}}\,.
\eea
We stress that non-linearity is the decisive obstacle here. Any linear functional would yield another Gaussian variable whose PDF can be derived from its 2-point function. That is, if the observables can be constructed from a set of linear combinations of the Gaussian field,
\be\label{eq:linear_G}
    \zeta_i \equiv \int \td^3 x \, L_i({\bf x}) \zeta_{\rm G}({\bf x})\,,
\ee
then their description reduces to a Gaussian one characterised by the covariance matrix $\langle\zeta_i \zeta_j\rangle$, which will completely determine the statistical features of the $\zeta$.

\subsection{Locally non-Gaussian fields}
\label{sec:loc_NG_basics}

The discussion above is very general. Therefore, we will restrict our attention to (i) \emph{locally non-Gaussian} fields that can be defined via
\be\label{eq:zetaNG}
     \zeta({\bf x}) = F[\zeta_{\rm G}({\bf x})]\,,
\ee
with $F$ a non-linear function that satisfies the property of (ii) being \emph{statistically homogeneous and isotropic} 
\be\label{eq:isohomo}
    \xi_{\rm G}({\bf x}_1,{\bf x}_2) = \xi_{\rm G}(|{\bf x}_1-{\bf x}_2|)\,.
\ee

Homogeneity and isotropy make it convenient to work in momentum space, where the 2-point function is described by the dimensionless power spectrum $\mathcal{P}_{\rm G}(k)$~\footnote{Whether we are working in momentum or position space is to be understood from the argument. The momentum and position space representations of the field are related through the Fourier transformation,
\be
    \zeta_{\rm G}({\bf k}) = \int \td^3 x \zeta_{\rm G}({\bf x}) e^{-i {\bf k} {\bf x}}, \qquad
    \zeta_{\rm G}({\bf x}) = \int \frac{\td^3 k}{(2\pi)^3} \zeta_{\rm G}({\bf k}) e^{i {\bf k} {\bf x}} \,.
\ee
These relations, as well as Eq.~\eqref{eq:xi_to_P} are also valid for non-Gaussian fields.}
\be\label{eq:PzetaG}
    \langle \zeta_{\rm G}({\bf k}) \zeta_{\rm G}({\bf k}')\rangle 
    = (2\pi)^3 \delta({\bf k}+{\bf k}') \frac{2\pi^2}{k^3}\mathcal{P}_{\rm G}(k)\,.
\ee
The power spectrum and the correlation function are related by
\be\label{eq:xi_to_P}
    \xi_{\rm G}(x) 
    = \int^{\infty}_0 \frac{\td k}{k}\, \frac{\sin(k x)}{k x}\mathcal{P}_{\rm G}(k),
    \qquad
    \mathcal{P}_{\rm G}(k) 
    = \frac{2k^2}{\pi}\int^{\infty}_0 \td x\, x \sin(k x) \xi_{\rm G}(x)\,.
\ee
The 2-point function \eqref{eq:PzetaG} is anti-diagonal in momentum space. The eigenfunctions of $\hat \xi_{\rm G}$ are $\zeta_{\rm G}({\bf k}) \pm \zeta_{\rm G}(-{\bf k})$, and it can be straightforwardly inverted so that, in momentum space, we have that 
\be
    \zeta_{\rm G} \hat\xi_{\rm G}^{-1} \zeta_{\rm G}
    = \int \frac{\td^3 k}{(2\pi)^3}  \frac{k^3 }{2\pi^2 \mathcal{P}_{\rm G}(k)} \zeta_{\rm G}({\bf k}) \zeta_{\rm G}(-{\bf k})\,.
\ee
This, together with the fact that spatial homogeneity forces the linear evolution of the field to be describable by multiplicative operators, makes it natural to describe curvature perturbations in momentum space at the linear level. Local non-linear modifications~\eqref{eq:zetaNG} will couple all scales, and thus it is generally necessary to resort to perturbation theory when estimating expectation values~\eqref{eq:n_point_general}. Evaluating the path integral~\eqref{eq:n_point_general} in momentum space offers a non-perturbative solution, but, in general, it can be computationally unfeasible. However, the computation can be significantly simplified in position space.

\paragraph{Determining the locality of non-Gaussian fields.} By Eq.~\eqref{eq:zetaNG} above, a locally non-Gaussian field can be defined via an auxiliary Gaussian field and the local mapping $F$. Such a definition of locality merely stipulates the existence of $\zeta_{\rm G}$ and $F$. However, it is not at all clear how to generally demonstrate that a non-Gaussian field is local if $\zeta_{\rm G}$ and $F$ are not given beforehand. Let us take a brief detour and consider this question in more detail before resuming the discussion of $N$-point functions.

In general, when $\zeta$ is \emph{locally} NG in the sense of Eq.~\eqref{eq:zetaNG}, there must exist an inverse function $F^{-1}$ that maps the non-Gaussian field into a Gaussian one as $\zeta_{\rm G}({\bf x}) = F^{-1}[\zeta({\bf x})]$, implying that such an approach only works when $F$ is invertible. Moreover, the following holds:
\begin{tcolorbox}
\emph{If $F$ is monotonous, it can be reconstructed from the 1-point distribution of $\zeta({\bf x})$.} 
\end{tcolorbox}

To see this, consider that when $\zeta({\bf x}) = F[\zeta_{\rm G}({\bf x})]$ and $F$ is monotonously growing, then the cumulative distribution functions $C(\zeta) \equiv P(\zeta({\bf x}) \leq \zeta)$ and $C_{\rm G}(\zeta_{\rm G}) = P_{\rm G}(\zeta_{\rm G}({\bf x}) \leq \zeta_{\rm G}) = [1+{\rm erf}(\zeta_{\rm G}/\sqrt{2\xi_0})]/2$ must satisfy,
\be\label{eq:F=CCG}
    C(F(\zeta_{\rm G})) = C_{\rm G}(\zeta_{\rm G})
    \qquad \Leftrightarrow \qquad
    F(\zeta_{\rm G})) = C^{-1}(C_{\rm G}(\zeta_{\rm G}))\,,
\ee
that is, $F$ is determined by the one-point distribution of the non-Gaussian field. Above, $\xi_0$ is the variance of the Gaussian field. Since $C(\zeta)$, $C_{\rm G}(\zeta_{\rm G})$ are monotonously growing, so is $F$. The case where $F$ is monotonously decreasing can be treated analogously by using $C_{\rm G}^{-1}(1-C(\zeta))$.

The situation is much more complicated when $F$ is not invertible, and thus there is no obvious way to go from $\zeta({\bf x})$ to $\zeta_{\rm G}({\bf x})$. Although the above construction \eqref{eq:F=CCG} can still be used to obtain a Gaussian one-point distribution by defining the monotonously increasing function $\bar F(\zeta_{\rm G}) \neq C^{-1}(C_{\rm G}(\zeta_{\rm G}))$. 
However, since $F(\zeta_{\rm G})$ is not monotonous, $F(\zeta_{\rm G}) \neq \bar F(\zeta_{\rm G})$ and $\bar F^{-1}(\zeta)$ will not yield the initial auxiliary Gaussian field. It is thus also not guaranteed that $\bar F^{-1}(\zeta)$ would be a Gaussian random field at all; that is, even if the $1$-point distribution is Gaussian, the $n$-point function might not be.

\subsection{A reduction of integration variables for local non-Gaussianity}

The computation of $n$-point functions of locally non-Gaussian fields is a balancing act between choosing the momentum representation, which significantly simplifies the linear theory, and the position representation in which the NG is quantified by the remarkably simple relation $\zeta({\bf x}) = F[\zeta_{\rm G}({\bf x})]$~\eqref{eq:zetaNG}. Below, we will first compute the position space $N$-point functions of $\zeta({\bf x})$ and then convert them to momentum space by evaluating their Fourier transform.

We need to estimate the expectation values for a Gaussian field $\zeta_{\rm G}({\bf x})$ at $n$ points ${\bf x}_i$. To shorten notation for later convenience, we will denote the Gaussian fields at ${\bf x}_i$ as 
\be
    \zeta_i \equiv \zeta_{\rm G}({\bf x}_i)
\ee
and the correlation function as\footnote{We drop the subindex ${G}$ denoting Gaussian variables only in the quantities $\zeta_i$, $\xi_{ij}$ and $\xi_{0}$.}
\be
    \xi_{ij} 
    \equiv \langle \zeta_i \zeta_j \rangle 
    \equiv \xi_{\rm G}(|{\bf x}_i - {\bf x}_j|), 
    \qquad
    \xi_{0} 
    \equiv \langle \zeta_i^2 \rangle
    \equiv \xi_{\rm G}(0)\,.
\ee
The $\zeta_i$ admit the form \eqref{eq:linear_G} with $L_i({\bf x}) = \delta({\bf x}_i - {\bf x})$. The distribution of $\zeta_i$ will be entirely determined by the finite covariance matrix $\xi_{ij}$ and the path integral  \eqref{eq:n_point_general} is reduced to a multi-dimensional Gaussian average. The $n$-point function is given by
\begin{tcolorbox}[ams align]
\label{eq:NG_NP_x}
    \left\langle\prod_i \zeta({\bf x}_i)\right\rangle
    = \frac{1}{\sqrt{\det (2\pi \xi_{ij})}} \int  \exp\left(- \frac{1}{2}\zeta_i (\xi^{-1})_{ij} \zeta_j  \right) \prod_i  F(\zeta_i) \td \zeta_i \equiv \mathcal{G}_{n}(\xi_{ij})\,,
\end{tcolorbox}
\noindent where $(\xi^{-1})_{ij}$ is the inverse covariance matrix. This simple but exact expression provides the foundation on which the rest of the analysis will be built.

The function $\mathcal{G}_{n}$ maps Gaussian $n$-point functions to non-Gaussian ones. It is a function of the covariance matrix or $1+(n-1)(n-2)/2$ variables $\xi_0$ and $\xi_{ij}$ with $1\leq i<j$. Importantly, it does not depend explicitly on coordinates; this dependence arises implicitly through $\xi_{ij}$. Therefore, 
\begin{tcolorbox}[ams align]
\emph{$\mathcal{G}_{n}$ is independent of the shape of the power spectrum}\nonumber
\end{tcolorbox}
\noindent and depends only on the integrated power of fluctuations $\xi_0 = \int \td \ln k\, \mathcal{P}_{\rm G}$. This is a considerable simplification, as it allows us to factorise the problem: the mapping $\mathcal{G}_{n}$ can be computed without fixing the power spectrum. Once $\mathcal{G}_{n}$ is known for a model $F$ of NG, any Gaussian power spectrum can be converted into a non-Gaussian $n$-point function
\be\label{eq:NG_NP_p}
    \left\langle \prod^{n}_{i=1} \zeta({\bf k}_i)\right\rangle' 
    = \int \mathcal{G}_{n}(\xi_{\rm G}(|{\bf x}_i-{\bf x}_j|))|_{{\bf x}_n = 0}\, \prod^{n-1}_{i=1}e^{-i {\bf x}_i {\bf k}_i}\, \td^3 x_i
    \,,
\ee
where the ' indicates that we impose ${\textstyle\sum}^{n}_{i=1} {\bf k}_i = 0$ and omit the $\delta$-function that arises due to homogeneity, that is, the full $n$-point function reads
\bea
    \left\langle \prod^{n}_{i=1} \zeta({\bf k}_i)\right\rangle 
    &= (2\pi)^3\delta\left(\sum^{n}_{i=1} {\bf k}_i\right) \left\langle \prod^{n}_{i=1} \zeta({\bf k}_i)\right\rangle'\,.
\eea
Together with the Fourier transformation, this amounts to $4n - 6$ integrals when $n>2$: $n$ when evaluating $\mathcal{G}_{n}$ and $3n$ when evaluating the Fourier transformation, which is reduced by $6$ when accounting for homogeneity and isotropy. For 2-point functions, the number of integrals is 3.

On the other hand, the number of terms grows exponentially with the perturbative order, and with it, the number of momentum integrals also increases. For non-Gaussian power spectra, that is, 2-point functions, the numerical evaluation of \eqref{eq:NG_NP_x} and \eqref{eq:NG_NP_p} can compete with perturbation theory even at the lowest order. For bi- and trispectra, numerical estimates based on conventional perturbation theory are expected to outperform the proposed non-perturbative approach at the first few orders.

\section{A semi-perturbative framework}
\label{sec:perturbations}

Aside from numerical estimates, position space $n$-point functions are also easier to study perturbatively. The perturbative approach is set up by expanding \eqref{eq:zetaNG} in powers of the Gaussian field
\be\label{eq:zeta_series}
    \zeta({\bf x}) = \sum^{\infty}_{n=1} F_{{\rm NL},n} (\zeta_{\rm G}^n({\bf x}) - \langle \zeta_{\rm G}^n({\bf x}) \rangle)\,,
\ee
where the subtracted expectation values ensure that $\langle\zeta({\bf x})\rangle = 0$. In perturbative estimates, we will follow convention and assume that $F_{{\rm NL},1} = F'[0] = 1$, so that $\zeta = \zeta_{\rm G}$ at the 0'th order. The next three indices are often denoted as $F_{{\rm NL},2} = (3/5) f_{\rm NL} \equiv F_{\rm NL}$, $F_{{\rm NL},3} = (3/5)^2 g_{\rm NL} \equiv G_{\rm NL}$, $F_{{\rm NL},4} = (3/5)^3 h_{\rm NL} \equiv H_{\rm NL}$. We will occasionally use the notation in capital letters to refer to lower-order perturbative corrections.

Below, we will derive the series expansion of $\mathcal{G}_{n}$ in $\xi_{ij}$ and show how each order is related to the series expansion~\eqref{eq:zeta_series} of $F$. We will show that $F_{{\rm NL},n}$ can be resummed at every order of $\xi_{ij}$ and that a perturbative approach does not rely on $F$ being analytic. The series expansion in $\xi_{ij}$ is not restricted to position space and can be straightforwardly converted to momentum space by replacing multiplication with convolutions; e.g., the term $\xi_{ij}^n$ is replaced by the $n$-th convolution power of~\eqref{eq:PzetaG}.

\subsection{Kibble–Slepian decomposition}

Any approach based on the series expansion~\eqref{eq:zeta_series} of $F[\zeta_{\rm G}]$ will yield a series expansion $\xi_{ij}$. Thus, it is natural to start with an expansion of $\xi_{ij}$ ($i\neq j$) of the multidimensional PDF in~\eqref{eq:NG_NP_x}. In the limit $\xi_{ij} \to \xi_0 \delta_{ij}$, i.e., at the 0-th order, the variables $\zeta_i$ become independent, and the expectation value factorises into vanishing one-point functions\footnote{At a single point,
\be
    \left\langle \zeta({\bf x}_1)\right\rangle 
    = \frac{1}{\sqrt{2\pi \xi_0}} \int \td \zeta_1\,  F(\zeta_1) \exp\left(- \frac{\zeta_1^2}{2 \xi_0} \right) \,.
\ee
Since $\left\langle \zeta({\bf x})\right\rangle$ would contribute to the background, not the curvature perturbations, we can take all 1P functions to vanish,
\be\label{eq:vanishing_mean}
    \left\langle \zeta({\bf x})\right\rangle \stackrel{!}{=} 0\, ,
\ee
without loss of generality. In case $\left\langle F(\zeta_{\rm G}) \right\rangle \neq 0$, we will make the shift $F(\zeta_{\rm G}) \to F(\zeta_{\rm G}) - \left\langle F(\zeta_{\rm G}) \right\rangle$.} and thus will also vanish itself.

The series expansion in correlation functions is obtained as a direct application of the Kibble–Slepian formula~\cite{Kibble1945,Slepian1972}. First, we must introduce some notation to formulate it. Consider a $n\times n$ symmetric matrix $\psi_{ij}$ with unit diagonal elements,
\be
    \psi_{ij}:  \quad 
    \psi_{ii} = 1\,, \quad
    \psi_{ij} = \psi_{ji}\,
\ee
and define the set $\mathcal{N}$ of $n\times n$ symmetric multiplicity matrices $\nu_{ij}$ with a vanishing diagonal ($\nu_{ii} = 0$) and with the off-diagonal entries being non-negative integers, such that
\be
    \mathcal{N} = \{ \nu_{ij} \in \mathbb{N}^{n\times n}: \nu_{ij} =  \nu_{ji}, \, \nu_{ii} = 0 \}\,.
\ee
For a given $\nu_{ij} \in \mathcal{N}$, we denote $s_{i} = \sum_j \nu_{ij}$. The Kibble–Slepian formula then states that
\be
    \frac{1}{\det(\psi_{ij})}\exp\left(\frac{1}{2}z_i z_i - \frac{1}{2} z_i (\psi^{-1})_{ij} z_j\right) = \sum_{\nu_{ij} \in \mathcal{N}} \tilde H_{s_1}(z_1) \ldots \tilde H_{s_n}(z_n) \prod_{i<j} \frac{\psi_{ij}^{\nu_{ij}}}{\nu_{ij}!}\,,
\ee
where $\tilde H_n(x) = 2^{-n/2} H_n(x/\sqrt{2})$ are rescaled versions of Hermite polynomials $H_n(x)$ and the sum runs over all multiplicity matrices in $\mathcal{N}$.

The Kibble–Slepian formula can be recast in a more suggestive form as a decomposition of the PDF of a $n$-dimensional Gaussian distribution,
\be
    p_n( \zeta_i, \xi_{ij}  ) 
    \equiv \frac{1}{\det(2\pi \xi_{ij})} \exp\left(- \frac{1}{2}\zeta_{i} (\xi^{-1})_{ij}\zeta_{j} \right)\,,
\ee
where $\zeta_i$ and $\xi_{ij}$ denote the Gaussian field at $n$ different points and the correlation matrix of these fields. Identifying $\psi_{ij} = \xi_{ij}/\xi_0$, $z_{i} = \zeta_{i}/\sqrt{\xi_0}$ gives
\be
    p_n( \zeta_i, \xi_{ij} ) 
    = \sum_{\nu_{ij} \in \mathcal{N}} \left[\prod_i  p_1( \zeta_{i}) \bar{H}_{s_i}(\zeta_i) \right] \prod_{i<j} \frac{\xi_{ij}^{\nu_{ij}}}{\nu_{ij}!}\,,
    \qquad
    \bar{H}_{s}(\zeta) \equiv \frac{1}{(2\xi_0)^{n/2}} H_n\left(\frac{\zeta}{\sqrt{2\xi_0}}\right)\,,
\ee
effectively decomposing the $n$-point distribution into a sum of distributions at a single point.  As a result, any $n$-point function \eqref{eq:NG_NP_x} can be expanded as
\begin{tcolorbox}[ams align]
\label{eq:NP_Cn_expansion}
    \left\langle\prod_i \zeta({\bf x}_i)\right\rangle
    = \sum_{\nu_{ij} \in \mathcal{N}}  \left[\prod_{i<j} \frac{ \xi_{ij}^{\nu_{ij}}}{\nu_{ij}!}\right] 
    \prod_i s_i!\mathcal{C}_{s_i}\,,
    \qquad\quad
    \mathcal{C}_{s} 
    \equiv \frac{1}{s!}\left\langle \bar{H}_{s}(\zeta_1) F[\zeta_1]\right\rangle
\end{tcolorbox}
\noindent yielding a closed-form expression for the perturbative expansion for $n$-point functions for arbitrary $n$, and it can be considered one of the main results of this study. The coefficients $\mathcal{C}_{s}$ are the only free parameters in the expansion and must encode all information about NG. They are given as an expectation value at a single point
\be\label{eq:Cs_integral}
    \mathcal{C}_{s} 
    = \frac{1}{s!}\int \frac{\td \zeta_{1}}{\sqrt{2\pi \xi_0}} \, \bar{H}_{s}(\zeta_{1}) F(\zeta_{1}) \exp(-\zeta_{1}^2/(2\xi_0))\,,
\ee
which can be accurately estimated numerically if an analytic approach is not feasible. Importantly, the coefficients $\mathcal{C}_s$ depend only on $\xi_0$ and the local mapping $F$, but not on the shape of the power spectrum. They are also \emph{uniquely determined} as long as the NG field $\zeta$ has finite variance (that is, $\langle F^2(\zeta_{\rm G})\rangle < \infty$), since Hermite polynomials constitute a complete orthogonal basis for the Hilbert space of functions that are square integrable with a Gaussian weight.\footnote{The rescaled Hermite polynomials $\bar H_m(\zeta_1)$ satisfy the orthogonality relation
\be
    \left\langle\bar{H}_{s}(\zeta_1)\bar{H}_{s'}(\zeta_1)\right\rangle 
    \equiv \int \frac{\td \zeta_1}{\sqrt{2\pi \xi_0}} \, \bar{H}_{s}(\zeta_1) \bar{H}_{s'}(\zeta_1) e^{-\frac{\zeta_1^2}{2\xi_0}} 
    = \frac{s!}{\xi_0^s} \delta_{ss'}\,.
\ee
} Consequently, the set of coefficients $\mathcal{C}_{s}$ specifies $F$ through its Hermite expansion
$F(\zeta_{1}) = \sum_s \mathcal{C}_{s} \,\xi_0^s\, \bar{H}_{s}(\zeta_{1})$.

Let us look at the expansion~\eqref{eq:NP_Cn_expansion} in more detail and begin by considering the relation~\eqref{eq:zeta_series} between $\zeta_{\rm G}$ and $\zeta$, which has been constructed in a way that makes the NG 1-point vanish. Therefore, since $\bar{H}_{0}(\zeta) = 1$, the 0th coefficient must vanish
\be
    \mathcal{C}_0 
    = \langle F[\zeta_1] \rangle = \langle \zeta({\bf x}_1) \rangle 
    = 0\,.
\ee
Thus, only terms with $s \geq 1$ contribute to the expansion~\eqref{eq:NP_Cn_expansion}. Additionally, for $s>0$, the shift $F[\zeta_1] \to F[\zeta_1] - \langle F[\zeta_1]\rangle$ can be omitted when evaluating the coefficients $\mathcal{C}_n$ because $\langle \bar H_{s}(\zeta_1) \rangle = 0$ when $s\geq1$.

To make contact with the $F_{{\rm NL},n}$ expansion~\eqref{eq:zeta_series}, we apply the identity 
\be
    \left\langle \zeta_1^n \bar H_m(\zeta_1) \right\rangle 
    = \begin{cases}
    \xi_0^{\frac{n-m}{2}} \frac{n!}{(n-m)!!}\, & n\geq m, n \equiv m \mod 2 \\
    0 &
    \end{cases}\,,
\ee
where !! denotes the double factorial. This gives
\be\label{eq:Cn_expansion}
    \mathcal{C}_s 
    = \sum_{m\geq 0} \frac{(s+2m)!}{s!(2m)!!} \xi_0^m F_{{\rm NL},s+2m} \,.
\ee
All the dependence on $F_{{\rm NL},n}$ has been absorbed into the $\mathcal{C}_s$ coefficients. This also holds for the dependence on $\xi_0$, which determines the Gaussian fluctuations in a point. From the perspective of perturbation theory, the Kibble–Slepian decomposition has \emph{resummed} all vertices due to fluctuations at a single point. We will demonstrate this explicitly in the next section~\ref{sec:diagrams} using diagrammatic techniques.

We included the factorial suppression when defining $\mathcal{C}_n$ in Eq.~\eqref{eq:2P_expansion_xi12}, so that the leading term of $\mathcal{C}_n$ is $F_{{\rm NL},n}$. Explicitly, the first four terms are
\be
\begin{array}{llllll}  
    \mathcal{C}_1 &= 1             &+ 3 \xi_0 G_{{\rm NL}}    &+ 15 \xi_0^2 F_{{\rm NL},5}  &+ 105 \xi_0^3 F_{{\rm NL},7} +\ldots\,, \\
    \mathcal{C}_2 &= F_{{\rm NL}}  &+ 6 \xi_0 H_{{\rm NL}}    &+ 45 \xi_0^2 F_{{\rm NL},6} &+ 420 \xi_0^3 F_{{\rm NL},8} +\ldots\,,  \\
    \mathcal{C}_3 &= G_{{\rm NL}}  &+ 10 \xi_0 F_{{\rm NL},5} &+ 105 \xi_0^2 F_{{\rm NL},7} &+ 1260 \xi_0^3 F_{{\rm NL},9} +\ldots\,, \\
    \mathcal{C}_4 &= H_{{\rm NL}}  &+ 15 \xi_0 F_{{\rm NL},6} &+ 210 \xi_0^2 F_{{\rm NL},8} &+ 3150 \xi_0^3 F_{{\rm NL},10} +\ldots\,.
\end{array}
\ee
Note that the combinatorial term in Eq.~\eqref{eq:2P_expansion_xi12} grows rapidly due to the double factorial in the denominator, so the convergence of the series is worse than the convergence of $F$ in the $F_{\rm NL,n}$ series. Estimating the $n$-point function by expanding it in $F_{\rm NL,n}$, as done in most of the literature, is thus likely to produce an asymptotic series that will fail at higher orders. An example of such a scenario will be given in Sec.~\ref{sec:examples}.

Since $F_{{\rm NL},n}$ appears in the $n$-point functions only inside $\mathcal{C}_s$, it is possible to work directly with $\mathcal{C}_s$. Importantly, $\mathcal{C}_s$ uniquely determines $F_{{\rm NL},n}$ at any given order via the inverse relation
\be\label{eq:Cn_inverse_expansion}
    F_{{\rm NL},n} = \sum_{m\geq 0} \frac{(n+2m)!}{n!(2m)!!} (-\xi_0)^m \mathcal{C}_{n+2m}\,.
\ee
Overall, as long as the perturbative approach is well behaved, there is no loss of information when swapping from $\mathcal{C}_s$ to $F_{{\rm NL},m}$ or back. Working at the $n$-th perturbative order, the first $\mathcal{C}_s$ coefficients map one-to-one to the first $n$ coefficients $F_{{\rm NL},m}$. Also, $F_{{\rm NL},m}$ does not depend on $\mathcal{C}_s$, when $s<m$ and vice versa. In particular, truncating either $\mathcal{C}_s$ at the order $s_{\rm cut}$, implies that $F_{{\rm NL}, s}$ are truncated at $s_{\rm cut}$, that is, 
\be 
    \mathcal{C}_s = 0, \quad s > s_{\rm cut}
    \qquad \Leftrightarrow \qquad
    F_{{\rm NL},s} = 0, \quad s > s_{\rm cut}\,.
\ee
We will therefore be working mostly with the coefficients $\mathcal{C}_s$, which can be considered more fundamental to the $n$-point functions than $F_{{\rm NL},s}$.

\subsection{A diagrammatic interpretation of the decomposition}
\label{sec:diagrams}

Diagrammatic techniques have proven to be an extremely powerful tool in perturbative field theory. It is therefore tempting to interpret the expansion \eqref{eq:NP_Cn_expansion} in terms of a diagrammatic formalism. Such techniques are not new in the context of locally non-Gaussian random fields and have been used for estimating corrections to SIGW backgrounds~\cite{Unal:2018yaa,Adshead:2021hnm,Perna:2024ehx,Li:2025met}.

Each term in the decomposition \eqref{eq:NP_Cn_expansion} can be represented by a diagram consisting of vertices labelled by $i$, lines between vertices labelled by $ij$, and the multiplicity of the lines given by $\nu_{ij}$. In this way, each diagram can be uniquely related to a matrix $\nu_{ij}$ over which the sum is taken. The basic elements of the perturbative expansion typically consist of vertices $F_{{\rm NL},n}$ and lines that are represented by correlation functions $\xi_{ij}$. However, \eqref{eq:NP_Cn_expansion} already reorganises the expansion and can be obtained from the following Feynman rules that consist of {\bf i)} lines with multiplicity $\nu_{ij}$,
\bea\label{eq:line_rule_X}
    \text{\includegraphics[width=0.65\textwidth]{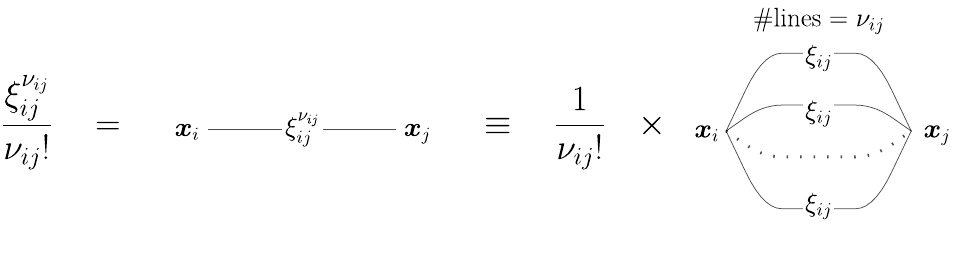}}
\eea
{\bf ii)} resummed vertices\footnote{Strictly speaking, “resummed” is a misnomer here, since the derivation of Eq.~\eqref{eq:NP_Cn_expansion} does not rely on a resummation procedure.}
\bea\label{eq:vertex_rule_X}
    \text{\includegraphics[width=0.85\textwidth]{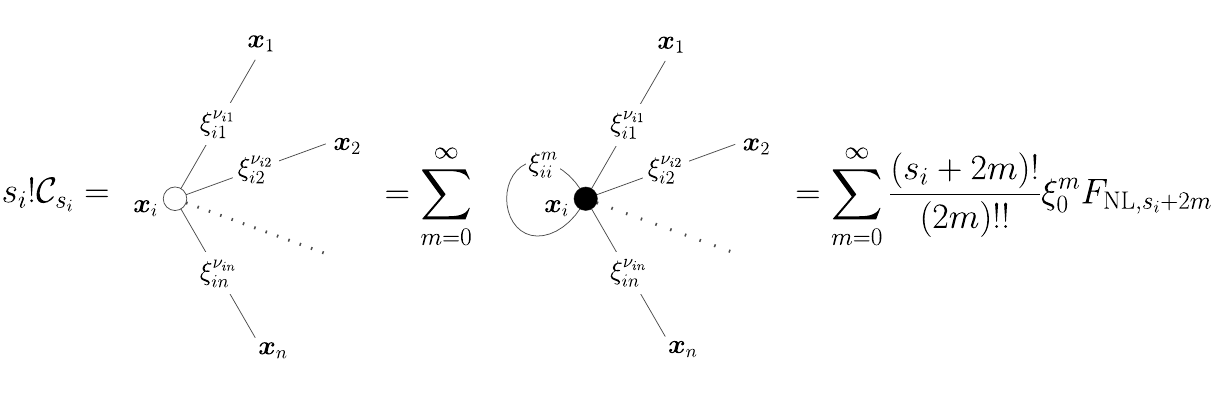}}
\eea
of order $s_i \equiv \sum^{n}_{j=1} \nu_{ij}$ that is given by the total multiplicity of the legs entering the vertex, and the condition that {\bf iii)} lines can only connect resummed vertices with different indices.

Let us try to understand the combinatorial factors and how these rules arise from the power series expansion \eqref{eq:NG_NP_p}. First, for the lines, the basic element is the line with multiplicity 1, which corresponds to $\xi_{ij}$. Lines with higher multiplicity arise when 2 points are connected by more than one correlation function, and the combinatorial factor in Eq.~\eqref{eq:line_rule_X} is a symmetry factor that avoids over-counting.

Second, a bare vertex with $n$ legs arises from the term $F_{{\rm NL},n} \zeta_{\rm G}({\bf x})^{n}$ in the expansion \eqref{eq:zeta_series}. Such vertices are represented by the black dot in \eqref{eq:vertex_rule_X}. As usual, correlation functions of Gaussian fields can be computed by Wick contracting all possible pairs of fields. 
Each contraction contributes a 2-point function. Thus, $2m$ of the fields in each term $F_{{\rm NL},n} \zeta_{\rm G}({\bf x})^{n}$ can be contracted with itself, and thus they would contribute to vertices that connect to $s = n - 2m$ other vertices. It follows that each vertex of order $s$ will receive contributions from terms $F_{{\rm NL},s+2m} \zeta_i^{s+2m}$, where $m \in \mathbb N$. The number of possibilities for connecting the $\zeta_i^{s+2m}$ to $s$ other vertices by contracting them with some $\zeta_j$ (with $j\neq i$) is $(s+2m)!/(2m)!$\,. On the other hand, the number of possible self contractions of the remaining $2m$ fields in $\zeta_i^{s+2m}$ is $(2m-1)!!$. The total multiplicity factor
\be
    \frac{(s+2m)!}{(2m)!} \times (2m-1)!! = \frac{(s+2m)!}{(2m)!!},
\ee
agrees with the one in Eq.~\eqref{eq:vertex_rule_X}. The $2m$ self contractions will additionally produce a factor of $\xi_{ii}^m \equiv \xi_0^m$. Combining all possible contributions will therefore yield the sum in \eqref{eq:vertex_rule_X}, which is identical to the expansion \eqref{eq:Cn_expansion} of $\mathcal{C}_s$ found using analytic techniques. We remark that the expansion of $\mathcal{C}_s$ in Eq.~\eqref{eq:Cn_expansion} agrees with the series expansions for resummed vertices obtained using an alternative derivation based on diagrammatic techniques for SIGW~\cite{Li:2025met}.

Although the diagrammatic approach using bare quantities agrees with the analytic one, we must stress that the integral formula for $\mathcal{C}_s$ in \eqref{eq:NP_Cn_expansion} is more general and can be applied even when $F(\zeta_{\rm G})$ does not admit a series expansion or at values of $\zeta_{\rm G}$ at which the series diverges. As a result, the diagrammatic formulation outlined in  Eq.~\eqref{eq:line_rule_X}, \eqref{eq:vertex_rule_X} formulated using the exact non-perturbative coefficients $\mathcal{C}_s$ can be applied if the expansion $F_{{\rm NL},n}$ fails.

The Feynman rules can be straightforwardly translated to $k$-space. Below we summarise the rules for the resummed perturbative formalism that yields $N$-point functions
\be
    \left\langle\prod_i \zeta({\bf x}_i)\right\rangle
    \qquad \mbox{and} \qquad
    \left\langle\prod_i \zeta({\bf k}_i)\right\rangle'\,.
\ee
They read:
\begin{enumerate}[label=\roman*.]
    \item Lines carry multiplicity and distance $(\nu_{ij}, x_{ij})$ in $x$-space or multiplicity and an internal momentum $(\nu_{ij}, {\bf q}_{ij})$ in $k$-space, with the momentum directed from $i \to j$ . Each line corresponds to
    \footnote{There is some abuse of terminology when we define the convolution  for dimensionless power spectra because they are not given purely as the Fourier transform of the correlation function. By accounting for the definition \eqref{eq:xi_to_P} the power spectrum corresponding to a correlation function $\xi(x) = \xi_a(x) \xi_b(x)$  given by the product of correlation functions $\xi_a(x)$ and $\xi_b(x)$ is
\bea\label{def:2P_convolutions}
    \mathcal{P}(k) 
    = (\mathcal{P}_a * \mathcal{P}_b)(k) 
    &\equiv \frac{k^3}{4\pi }\int \frac{\td^3 q}{q^3 |{\bf k-q}|^3} \mathcal{P}_a(q) \mathcal{P}_b(|{\bf k-q}|)\,,
\eea
where $\mathcal{P}_a$, $\mathcal{P}_b$ are power spectra corresponding to $\xi_a(x)$ and $\xi_b(x)$, respectively. This defines the convolution of dimensionless power spectra used in Eq.~\eqref{eq:Feyn_line}. By making use of isotropy, the convolution integral can be recast as
\bea
    (\mathcal{P}_a * \mathcal{P}_b)(k) 
    &= \frac{k^2}{2} \int_{\mathcal{D}_k}\frac{\td q_1 \td q_2}{q_1^2 q_2^2}\mathcal{P}_b\left(q_1\right) \mathcal{P}_a\left(q_2\right)\,,
    \\
    &= 4 \int^{\infty}_1 \td t \int^1_{-1} \td s \frac{1}{(t^2-s^2)^2}\mathcal{P}_b\left(k \frac{t+s}{2}\right) \mathcal{P}_a\left(k \frac{t-s}{2}\right)\,,
\eea
where the domain of integration $\mathcal{D}_k$ in the first expression is given by $q_1,q_2>0$ and $|q_1-q_2|\leq k\leq q_1+q_2$.The integration variables $s$ and $t$ are analogous to those that often appear when computing scalar-induced GW signals (see e.g.~\cite{Vaskonen:2020lbd, Franciolini:2023pbf}).}  
    \be\label{eq:Feyn_line}
        \text{\includegraphics[width=0.2\linewidth, valign=c]{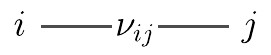} :}
        \qquad
        \xi_{\rm G}^{\nu_{ij}}(x_{ij})
        \quad \mbox{or} \quad 
        P_{\rm G}^{*\nu_{ij}}(q_{ij}) \equiv \frac{2\pi^2}{q_{ij}^3} \mathcal{P}_{\rm G}^{*\nu_{ij}}(q_{ij})\,,
    \ee
    where $P_{\rm G}^{*n}$ denotes the $n$-th convolution power of the Gaussian power spectrum $P_{\rm G}$ and $\mathcal{P}_{\rm G}^{*\nu_{ij}}(q_{ij})$ is the dimensionless power spectrum corresponding to $\xi_{\rm G}^{\nu_{ij}}(x_{ij})$.
       
    \item Vertices carry multiplicity and position $(s_{i}, {\bf x}_{i})$ in $x$-space or multiplicity and an external momentum $(s_{i}, {\bf k}_{i})$) in $k$-space. In both cases, the Feynman rule is the same
    \be\label{eq:Feyn_vertex}
        \text{\includegraphics[width=0.24
        \linewidth, valign=c]{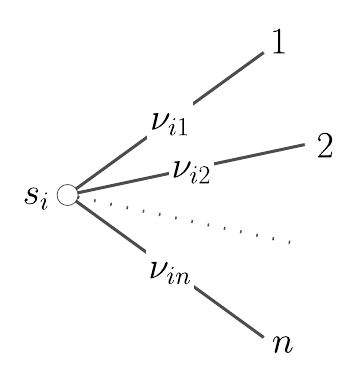} :}
        \qquad \qquad
        s_i! \mathcal{C}_{s_i}\,.
    \ee
    
    \item The multiplicities satisfy: a) Multiplicities are not directional $\nu_{ij} = \nu_{ji} \in \mathbb N$. b) The multiplicity of lines connecting the same vertex vanishes $\nu_{ii} = 0$, so such lines are not allowed. c) Multiplicities are "conserved" at each vertex
    \be\label{eq:order_c}
        s_i = \sum^{n}_{j=1} \nu_{ij}\,.
    \ee
    
    \item[iv.($k$)] In $k$-space, momenta are conserved at each vertex 
    \be\label{eq:momentum_c}
        {\bf k}_{i} = \sum^{n}_{j=1} q_{ij}\,.
    \ee
    Each unconstrained internal momentum (loop) must be integrated over $\int \td q_{ij}^3/(2\pi)^3$. The momenta (not explicitly shown above) can be taken to satisfy ${\bf q}_{ij} = - {\bf q}_{ji}$ to simplify bookkeeping.

    \item[iv.($x$)] The analogue of the last rule is, that, in $x$-space, the lines carry the distance between the vertices, $x_{ij} = |{\bf x}_{i} - {\bf x}_{j}|$.
\end{enumerate}

\subsection{2-point functions and non-Gaussian power spectra}

Let us consider the non-linear 2-point function and the corresponding power spectrum. In this case, determining the non-linear power spectrum has been reduced to computing the function
\be\label{eq:G2_bounds}
    \mathcal{G}_2: [-\xi_0, \xi_0] \to [-\mathcal{G}_2(\xi_{0}), \mathcal{G}_2(\xi_{0})]\,,
\ee
that, by Eq.~\eqref{eq:NG_NP_x}, can be expressed by the double integral\footnote{We keep the dependence on $\xi_0$ in $\mathcal{G}_2$ implicit.} 
\bea\label{eq:NG_2P_function}
    \mathcal{G}_2(\xi_{12}) \equiv \left\langle F(\zeta_1) F(\zeta_2) \right\rangle 
    &=  \int \frac{\td \zeta_1 \td \zeta_2\,  F(\zeta_1) F(\zeta_2)}{2\pi \sqrt{\xi_0^2 - \xi_{12}^2}} e^{- \frac{1}{2} \frac{(\zeta_1^2 + \zeta_2^2) \xi_0  - 2\zeta_1 \zeta_2 \xi_{12} }{\xi_0^2 - \xi_{12}^2} } \,.
\eea
Explicitly, expressed in terms of the Gaussian correlation function $\xi_{\rm G}(x)$, the non-linear correlation function is 
\be
    \xi(x) \equiv \langle \zeta({\bf x}) \zeta(0)\rangle = \mathcal{G}_2(\xi_{\rm G}(x))
\ee
and the corresponding non-linear power spectrum is
\be\label{eq:PS_NG}
    \mathcal{P}(k) 
    = \frac{2k^2}{\pi}\int^{\infty}_0 \td x\, x \sin(k x)\mathcal{G}_2(\xi_{\rm G}(x))\,.
\ee

Some of the basic properties of $\mathcal{G}_2$ can be laid out in general, without reference to a specific realisation of $\zeta = F(\zeta_{\rm G})$. When the Gaussian field is uncorrelated, the expectation value factorises $\langle F(\zeta_1) F(\zeta_2) \rangle = \langle F(\zeta_1)\rangle \langle F(\zeta_2) \rangle = 0$ and vanishes because $\langle F(\zeta_1)\rangle = 0$ by construction, so that 
\be
    \mathcal{G}_2(0) = 0\,
\ee
and uncorrelated points are therefore always mapped to uncorrelated ones. In the maximally (anti)correlated case when $\xi_{12} = (-)\xi_0$, we must have $\zeta_1 = (-)\zeta_2$ and the double integral~\eqref{eq:NG_2P_function} is reduced to a single integral
\be
    \mathcal{G}_2(\xi_{0}) 
    = \left\langle F(\zeta_1)^2 \right\rangle\,,
    \qquad
    \mathcal{G}_2(-\xi_{0}) = \left\langle F(\zeta_1) F(-\zeta_1) \right\rangle\,.
\ee
The Cauchy-Schwartz inequality gives $|\mathcal{G}_2(\xi_{12})| \leq \mathcal{G}_2(\xi_{0})$ for all $\xi_{12} \in [-\xi_0,\xi_0]$ which corresponds to the range reported in Eq.~\eqref{eq:G2_bounds}. As maximally correlated points are mapped to maximally correlated ones, then inequality is saturated when $\xi_{12} = \xi_{0}$.  This does, however, not hold in the maximally anti-correlated Gaussian case, as anti-correlated points in the Gaussian field do not have to translate into anti-correlated points in the non-Gaussian field. A simple example is given by $\zeta({\bf x}) \propto \zeta_{\rm G}^2({\bf x}) - \langle \zeta_{\rm G}^2({\bf x}) \rangle$, that gives $\mathcal{G}_2(\xi_{12}) \propto \xi_{12}^2$ so all points of the non-Gaussian field $\zeta({\bf x})$ are positively correlated ($\xi(x) > 0$) independently of the shape of the Gaussian correlation function.

\begin{figure}[t]
\centering
\includegraphics[width=0.25\textwidth]{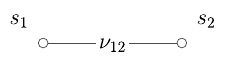}
\caption{Diagrams contributing to the correlation function and the power spectrum. The line and the vertices carry the same multiplicity $n \equiv \nu_{12}=s_1=s_2 \in \mathbb{N}$ and the same momentum. } 
\label{fig:2P}
\end{figure}

In general, the integral \eqref{eq:NG_2P_function} does not admit an analytic form, but it can be evaluated numerically (see Sec.~\ref {sec:examples}). However, for a near-Gaussian field $\zeta$, it can be approximated perturbatively via the expansion \eqref{eq:NP_Cn_expansion}. The multiplicity matrices for the bivariate distribution are
\be\label{eq:2P_N_matrix}
    \mathcal{N} = \left\{
    \begin{pmatrix}
        0 & n \\
        n & 0
    \end{pmatrix}: n \in \mathbb{N}_{+}
    \right\}\,,
\ee
so that each multiplicity matrix/diagram is completely quantified by a single integer, $n \equiv \nu_{12} = s_1 = s_2$. The Kibble-Slepian formula reduces to Mehler's formula 
\be
    \frac{p_2(\zeta_1,\zeta_2;\xi_{12},\xi_0)}{p_2(\zeta_1)p_2(\zeta_2)}
    = \sum^{\infty}_{n=1} \frac{\xi_{12}^n}{n!} \bar{H}_n(\zeta_1) \bar{H}_n(\zeta_2)\,,
\ee
and the non-Gaussian 2-point function reduces to a power series of Gaussian 2-point functions,
\be\label{eq:2P_expansion_xi12}
    \mathcal{G}_2 (\xi_{12})
    = \sum^{\infty}_{n=1} n! \mathcal{C}_n^2 \xi_{12}^n \,.
\ee
The corresponding diagrams are given in Fig.~\ref{fig:2P}.

In momentum space, the powers of $\xi_{12}^n$ are replaced by convolution powers of the Gaussian power spectrum. Either by taking the Fourier of \eqref{eq:2P_expansion_xi12} or by following the $k$-space Feynman rules for the series diagrams in Fig.~\ref{fig:2P}, the expansion \eqref{eq:2P_expansion_xi12} can be recast as
\be\label{eq:2P_expansion_PG}
    \mathcal{P}(k) 
    = \sum^{\infty}_{n=1} n! \mathcal{C}_n^2 \mathcal{P}_{\rm G}^{*n}(k)\,,
\ee
where $\mathcal{P}_{\rm G}^{*n}$ denotes the $n$-th convolution power of the Gaussian power spectrum $\mathcal{P}_{\rm G}$.

We remark that, by Eq.~\eqref{def:2P_convolutions}, the convolutions tend to possess a $k^3$ "causality" tail in the IR, e.g.,
$
    \mathcal{P}_{\rm G}^{*2}(k)  \sim k^3 \int \mathcal{P}_{\rm G}^{2}(p)\,\td p/p^4
$. 
If $\mathcal{P}_{\rm G}(k) \propto k^{n}$, with $n<-3$, this tail dominates the non-Gaussian power spectrum if $C_n$ are sufficiently large. Indeed, from \eqref{eq:PS_NG}, we find in the $k\to0$ limit that
\be\label{eq:k3_tail}
    \mathcal{P}(k) 
    \stackrel{k \ll k_{\rm IR}}{\sim} \mathcal{G}_2(\xi_0) \left(\frac{k}{k_{\rm IR}}\right)^3\,,
    \qquad
    k_{\rm IR} \equiv \left(\frac{2}{\pi}\int^{\infty}_0 \td x\, x^2 \frac{\mathcal{G}_2(\xi_{\rm G}(x))}{\mathcal{G}_2(\xi_0)}\right)^{-1/3}\,,
\ee
where the IR scale $k_{\rm IR}$ is determined by the average volume within which the field remains correlated (note that $\mathcal{G}_2(\xi_{\rm G}(x))/\mathcal{G}_2(\xi_0) = \xi(x)/\xi(0)$ gives the correlation of the non-Gaussian field.) The $k^3$ scaling is expected to hold when $k \ll k_{\rm IR}$. 
Spectra with a faster than $k^3$ IR-growth correspond to cases in which $k_{\rm IR}$ diverges. This requires a specific shape of $\mathcal{G}_2(\xi_{\rm G}(x))$, which is unlikely to occur if $\mathcal{G}_2$ and $\xi_{\rm G}(x)$ are unrelated. As a result, the $k^3$ IR-tails are expected to be a fairly general feature of locally non-Gaussian fields.

\subsection{Bispectra} 

The bispectrum corresponds to the connected 3-point correlation function, and thus, by~\eqref{eq:NG_NP_p}, it can be constructed from the Gaussian 2-point function as
\bea\label{eq:bispectrum}
    B({\bf k}_1,{\bf k}_2) 
    &\equiv \langle \zeta({\bf k}_1)\zeta({\bf k}_2)\zeta({\bf k}_3) \rangle' \,
    \\
    &= \int \td x^3_1 \td x^3_2 \, e^{-i ({\bf x}_1 {\bf k}_1 + {\bf x}_2 {\bf k}_2)}  \mathcal{G}_3(\xi_{\rm G}( x_1),\xi_{\rm G}(x_2),\xi_{\rm G}(|{\bf x}_1 -{\bf x}_2|))\,,
\eea
where ${\bf k}_1+{\bf k}_2+{\bf k}_3=0$. Since the one-point function vanishes, there are no disconnected contributions, making the bispectrum the leading pure probe of NG.  Statistical isotropy allows us to express the bispectrum as a function of the two moduli $k_1$, $k_2$ and an additional angle $\angle({\bf k}_1,{\bf k}_2)$ or moduli $k_3 \equiv |{\bf k}_1 - {\bf k}_2|$ so that the bispectrum can be expressed either as a function of two momenta $B({\bf k}_1,{\bf k}_2)$ or three moduli $B(k_1,k_2,k_3)$. The mapping $\mathcal{G}_3(\xi_{12},\xi_{13},\xi_{23})$ is non-perturbatively given by the 3-dimensional integral~\eqref{eq:NG_NP_x}.

\begin{figure}[t]
\centering
\includegraphics[width=0.45\textwidth]{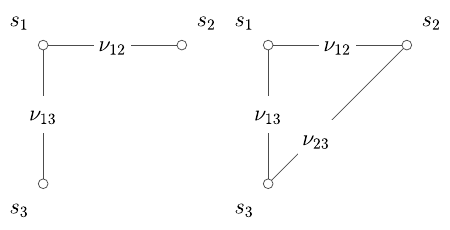}
\caption{Diagrams contributing to the 3-point function up to permutations of the vertices. Momenta (or positions) at the vertices and lines are not shown.} 
\label{fig:3P_diagrams}
\end{figure}

Perturbatively, the 3-point function can be split into terms that depend on products of 2 or 3 different correlation functions, as depicted diagrammatically in Fig.~\ref{fig:3P_diagrams}, modulo vertex permutations. Thus, the 3-point function can be expressed as
\be\label{eq:G3}
    \mathcal{G}_3(\xi_{12},\xi_{13},\xi_{23}) 
    =
    \mathcal{G}^{(2)}_3(\xi_{12},\xi_{13}) + \mathcal{G}^{(2)}_3(\xi_{12},\xi_{23}) + \mathcal{G}^{(2)}_3(\xi_{23},\xi_{13}) + \mathcal{G}^{(3)}_3(\xi_{12},\xi_{13},\xi_{23})\,,
\ee
where
\bea
    \mathcal{G}^{(2)}_3(\xi_{12},\xi_{13}) 
    &= \sum_{s_2,s_3\geq 1}  (s_2+s_3)!
    \mathcal{C}_{s_2} \mathcal{C}_{s_3} \mathcal{C}_{s_2+s_3} \xi_{12}^{s_2}\xi_{13}^{s_3}\,,
    \\
    \mathcal{G}^{(3)}_3(\xi_{12},\xi_{13},\xi_{23})
    &= \sum_{\nu_{12},\nu_{13},\nu_{23}\geq1} \frac{s_1!s_2!s_3!}{\nu_{12}!\nu_{13}!\nu_{23}!}
    \mathcal{C}_{s_1} \mathcal{C}_{s_2} \mathcal{C}_{s_3} \xi_{23}^{\nu_{23}}\xi_{13}^{\nu_{13}}\xi_{12}^{\nu_{23}}\,
\eea
are obtained from diagrams that contain 2 and 3 legs, respectively. In the 2-legged diagram, multiplicity conservation Eq.~\eqref{eq:order_c} gives $\nu_{13}=s_3$, $\nu_{12}=s_2$, so that $s_1 = s_2 + s_3$. For the 3-legged diagrams, we have that $s_1 = \nu_{12} + \nu_{13}$, $s_2 = \nu_{12} + \nu_{23}$, $s_3 = \nu_{13} + \nu_{23}$.

The bispectrum inherits the decomposition in \eqref{eq:G3} into terms depending explicitly on 2 or 3 momenta 
\be
    B(k_1,k_2,k_3) 
    = B^{(2)}(k_1,k_2) + B^{(2)}(k_1,k_3) + B^{(2)}(k_2,k_3) + B^{(3)}(k_1,k_2,k_3)\,.
\ee
The $B^{(2)}$ terms do not contain any loops and thus it reduces to a sum of products of the convolution powers of the power spectrum
\bea\label{eq:B_2}
    B^{(2)}(k_2,k_3) 
    &= \frac{(2\pi^2)^2}{k_2^3 k_3^3}\sum_{s_2,s_3 \geq 1}  (s_2+s_3)!
    \mathcal{C}_{s_2} \mathcal{C}_{s_3} \mathcal{C}_{s_2+s_3} \mathcal{P}_{\rm G}^{*s_2}(k_2)\mathcal{P}_{\rm G}^{*s_3}(k_3)
\eea
because one can always eliminate the third coordinate, e.g., $\xi_{13}^n \xi_{23}^m \to \xi^n_{\rm G}(x_1)\xi^m_{\rm G}(x_2)$ and then use that the Fourier transform of $\xi^n_{\rm G}(x)$ is $(2\pi)^2 \mathcal{P}_{\rm G}^{*n}(k)/k^3$, as in the case of the 2-point functions. This is not the case for the diagrams with 3 legs, for which we obtain
\bea\label{eq:B_3}
    B^{(3)}(k_1,k_2,k_3) 
    &= \sum_{\nu_{12},\nu_{13},\nu_{23}\geq 1} \frac{s_1!s_2!s_3!}{\nu_{12}!\nu_{13}!\nu_{23}!}
    \mathcal{C}_{s_1} \mathcal{C}_{s_2} \mathcal{C}_{s_3}
    \\
    &\times \int \frac{\td^3 q}{(2\pi)^3}
    P_{\rm G}^{*\nu_{12}}(q)   
    P_{\rm G}^{*\nu_{13}}({\bf q} - {\bf k}_1)
    P_{\rm G}^{*\nu_{23}}({\bf q} + {\bf k}_2)\,.
\eea

The $B^{(3)}$ term contributes only at the next-to-leading order. The sum in~\eqref{eq:B_3} runs over strictly positive $\nu_{ij} \geq 1$, such that $s_i \geq 1$, and thus the lowest order term is proportional to $\mathcal{C}_2^3$ (or $F_{\rm NL,2}^3$), while the leading term in~\eqref{eq:B_2} appears at the $\mathcal{C}_1^2 \mathcal{C}_2$ (or $F_{\rm NL,2}$) order. Otherwise, Eq.~\eqref{eq:B_2} can be absorbed into Eq.~\eqref{eq:B_3} by including terms with $\nu_{ij} = 0$.

\subsection{4P: Trispectra}

\begin{figure}[t]
\centering
\includegraphics[width=0.7\textwidth]{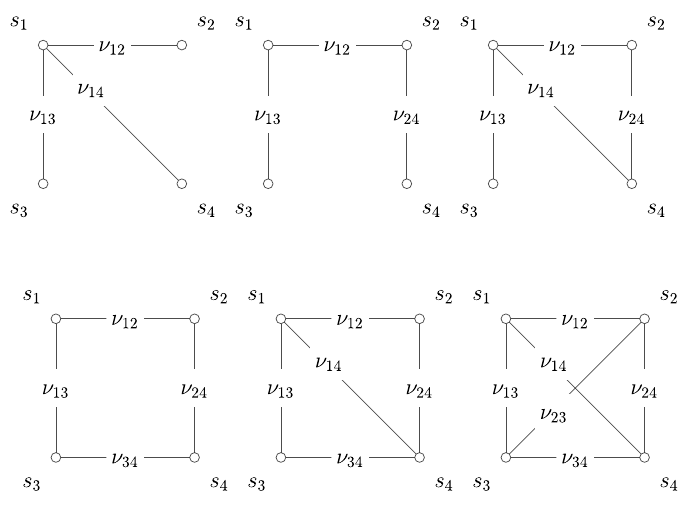}
\vspace{-3mm}
\caption{Diagrams contributing to the 4-point function up to permutations of the vertices.} 
\label{fig:4P_diagrams}
\end{figure}

The trispectrum is the connected component of the 4-point function. In terms of multiplicity matrices, the three terms of the disconnected component arise from the subset
\be
    \mathcal{N}_{\rm dis} 
    = \left\{
    \begin{pmatrix}
        0 & n & 0 & 0\\
        n & 0 & 0 & 0\\
        0 & 0 & 0 & m\\
        0 & 0 & m & 0
    \end{pmatrix}, 
    \begin{pmatrix}
        0 & 0 & n & 0\\
        0 & 0 & 0 & m\\
        n & 0 & 0 & 0\\
        0 & m & 0 & 0
    \end{pmatrix},
    \begin{pmatrix}
        0 & 0 & 0 & n\\
        0 & 0 & m & 0\\
        0 & m & 0 & 0\\
        n & 0 & 0 & 0
    \end{pmatrix}
    : n,m \in \mathbb{N}_{+}
    \right\} \subset \mathcal{N}\,,
\ee
while the rest $\mathcal{N}_{\rm con} = \mathcal{N} \setminus \mathcal{N}_{\rm dis}$ give rise to the connected component. In this way, we obtain the usual decomposition into connected and disconnected components,
\bea
    \langle \zeta({\bf x}_1) \zeta({\bf x}_2) \zeta({\bf x}_3) \zeta({\bf x}_4)\rangle' 
    &= 
    \mathcal{G}_2(\xi_{12}) \mathcal{G}_2(\xi_{34}) + 
    \mathcal{G}_2(\xi_{13}) \mathcal{G}_2(\xi_{24}) +
    \mathcal{G}_2(\xi_{14}) \mathcal{G}_2(\xi_{23})\\
    &+
    \mathcal{G}_{4,\rm con}(\xi_{12}, \xi_{13},\xi_{14}, \xi_{23},\xi_{24},\xi_{34})\,,
\eea
via perturbation theory. The connected component $\mathcal{G}_{4,\rm con}$ is a function of 7 variables ($\xi_{ij}$ where $i>j$ and $\xi_0$) that maps the Gaussian 2-point functions into the connected 4-point function of the non-Gaussian field $\zeta$. 

The Feynman diagrams contributing to the 4P functions are listed in Fig.~\ref{fig:4P_diagrams}, up to permutations. Analogously to the case of the 3-point function, the first 5 graphs can be obtained from the last by setting some $\nu_{ij}$ to 0. Explicitly, the 4-point function is given by the series of 6 variables,
\vspace{-1mm}
\bea\label{eq:4P_expansion}
    \mathcal{G}_{4,\rm con}(\xi_{12}, &\xi_{13},\xi_{14}, \xi_{23},\xi_{24},\xi_{34})
    = \!\!\!\! \sum_{\nu_{ij} \in \mathcal{N}_{\rm con} } \!\!
    \frac{s_1!s_2!s_3!s_4! \mathcal{C}_{s_1} \mathcal{C}_{s_2} \mathcal{C}_{s_3} \mathcal{C}_{s_4}}{\nu_{12}!\nu_{13}!\nu_{14}!\nu_{23}!\nu_{24}!\nu_{34}!}
    \xi_{12}^{\nu_{12}}
    \xi_{13}^{\nu_{13}}
    \xi_{14}^{\nu_{14}}
    \xi_{23}^{\nu_{23}}
    \xi_{24}^{\nu_{24}}
    \xi_{34}^{\nu_{34}}
    \\ 
    &=  
    (2!)^2\mathcal{C}_1^2\mathcal{C}_2^2  
    \sum_{\sigma}\xi_{\sigma_1 \sigma_2} \xi_{\sigma_1 \sigma_3} \xi_{\sigma_2 \sigma_4} 
    \\
    &+ 3! \mathcal{C}_1^3\mathcal{C}_3
    \sum_{\sigma}\xi_{\sigma_1 \sigma_2} \xi_{\sigma_1 \sigma_3} \xi_{\sigma_1 \sigma_4} 
    \\
    &+ (2!)^4 \mathcal{C}_2^4 \sum_{\sigma}\xi_{\sigma_1 \sigma_2} \xi_{\sigma_2 \sigma_3}\xi_{\sigma_3 \sigma_4} \xi_{\sigma_1 \sigma_4} 
    \\
    &+ \frac{(3!)^2}{2} \mathcal{C}_1^2\mathcal{C}_3^2 \sum_{\sigma}\xi_{\sigma_1 \sigma_2}^2 \xi_{\sigma_1 \sigma_3}\xi_{\sigma_2 \sigma_4}
    \\
    &+ (2!)^2 3! \mathcal{C}_1 \mathcal{C}_2^2 \mathcal{C}_3  \bigg[ 
    \frac{1}{2}\sum_{\sigma}\xi_{\sigma_1 \sigma_2} \xi_{\sigma_1 \sigma_3}^2\xi_{\sigma_3 \sigma_4}
    +
    \sum_{\sigma}\xi_{\sigma_1 \sigma_2} \xi_{\sigma_1 \sigma_3}\xi_{\sigma_1 \sigma_4}\xi_{\sigma_2 \sigma_3}
    \bigg]
    \\
    &+ 4! \mathcal{C}_1^2\mathcal{C}_2\mathcal{C}_4 
    \sum_{\sigma}\xi_{\sigma_1 \sigma_2}^2 \xi_{\sigma_1 \sigma_3} \xi_{\sigma_1 \sigma_4} 
    + \ldots\,,
\eea
where $s_i = \sum^{4}_{j = 1} \nu_{ij}$. Let us define the multiplicity\footnote{The lowest order (Gaussian) disconnected term of the 4-point function has $\nu_{\rm tot} = 2$. Thus, the order can be defined as $\nu_{\rm tot} - 2$. We will not introduce this notion, as it cannot be unambiguously generalized to $n$-point functions with odd $n$.}

\be
    \nu_{\rm tot} \equiv \sum_{i<j} \nu_{ij} = \sum_i s_i/2\,.
\ee
It counts the number of Gaussian correlation functions/power spectra contributing to a given term in the expansion. The explicit sum in Eq.~\eqref{eq:4P_expansion} shows all terms up to multiplicity $\nu_{\rm tot} \leq 4$: the first two rows have $\nu_{\rm tot} = 3$, and the remaining four have $\nu_{\rm tot} = 4$. Only the first 4 graphs appear, as the last two have 5 and 6 legs, respectively, and the number of legs cannot be smaller than $\nu_{\rm tot}$ by construction.

In momentum space, the first two graphs, containing 3 lines, generate the lowest order contributions and can be expressed as products of cumulant powers of Gaussian power spectra. The remaining diagrams will contain at least one loop integral. In general, diagrams with $n$ lines and $m$ vertices will have $n-m+1$ unconstrained momenta and thus $n-m+1$ loop integrals.

\subsection{Counting diagrams}

When applying the expansion~\eqref{eq:NP_Cn_expansion}, it is also useful to estimate the growth of the number of terms (or diagrams) at a given order. Since every diagram corresponds to a multiplicity matrix $\nu_{ij}$, it suffices to count multiplicity matrices for a fixed $\nu_{\rm tot}$.  For an $N$-point function the multiplicity matrix contains $M = N(N-1)/2$ independent elements. The problem therefore reduces to counting the number of ways $M$ non-negative integers can sum to $\nu_{\rm tot}$, with the constraint that every vertex should have multiplicity $s_i \ge 1$ (since $\mathcal{C}_0 = 0$). This implies that $\nu_{\rm tot} \geq N/2$ as each vertex must be connected to at least one line. Assuming connectedness would further require that $\nu_{\rm tot} \geq N-1$. Ignoring such constraints for the moment, the number of diagrams contributing to an $N$-point function at order $\nu_{\rm tot}$ is
\be
    \frac{[\nu_{\rm tot}+M-1]!}{\nu_{\rm tot}!\,(M-1)!},
\ee
from which one must subtract diagrams containing vertices with zero multiplicity. Thus, when $\nu_{\rm tot} \gg 1$, the number of diagrams grows polynomially as
\be
    \#\mbox{diagrams} \sim \frac{\nu_{\rm tot}^{M-1}}{(M-1)!}.
\ee
For the three cases considered above this implies a $\nu_{\rm tot}^0$ growth when $N=2$, consistent with a single diagram at each order, a $\nu_{\rm tot}^2$ growth when $N=3$, and a $\nu_{\rm tot}^5$ growth when $N=4$. Importantly, the number of diagrams grows \emph{polynomially} in the total multiplicity but \emph{exponentially} with the number of fields entering the correlation function. This is evident from the large number of terms contributing to the connected 4-point function at the leading ($\nu_{\rm tot} =3$) an next-to-leading ($\nu_{\rm tot} =4$) orders in Eq.~\eqref{eq:4P_expansion}.

This scaling behaviour is not affected by subtracting diagrams containing vertices with zero multiplicity, since their number grows slower than the total number of diagrams. The same holds for excluding disconnected diagrams. Counting diagrams only up to permutations of the vertices merely reduces the total number by a combinatorial factor, but does not change the exponent of polynomial growth in $\nu_{\rm tot}$.

The rapid growth of the number of diagrams translates into a comparable growth in the number of integrals that must be evaluated in order to estimate polyspectra at a high order. Thus there must be an order at which the numerical estimation of the fully non-perturbative polyspectra~\eqref{eq:NG_NP_p},~\eqref{eq:NG_NP_x} becomes computationally more efficient than the series expansion~\eqref{eq:NP_Cn_expansion}. However, when $N\geq 3$, the dimensionality of the integrals in the fully non-perturbative~\eqref{eq:NG_NP_x} is generally larger than the dimensionality of the integrals involved in the series, it is not trivial to estimate the order at which~\eqref{eq:NG_NP_p},~\eqref{eq:NG_NP_x} will give a computational advantage over the series~\eqref{eq:NP_Cn_expansion}.

\section{Exponentially tailed locally non-Gaussian fields}
\label{sec:examples}

Whenever the non-Gaussian field is constructed from the Gaussian one via a polynomial, an exact description is possible through perturbative methods. The above formalism allows us to look not just beyond polynomial relations~\eqref{eq:zetaNG}, but also beyond cases where a series expansion is not possible, e.g., when the radius of convergence is small or when $F$ is a piecewise function. This is also possible in the resummed perturbative formulation in terms of $\mathcal{C}_s$ as the integral~\eqref{eq:Cs_integral} does not require $F$ to be analytic. 

As an example, we will consider models that exhibit exponential tails in the 1-point distribution, for which
\bea\label{eq:F_USR_abs}
    \zeta({\bf x}) 
    &= - \frac{1}{2\beta}\ln\left|1-2\beta \zeta_{\rm G}({\bf x})\right| \\
    &= \sum^{\infty}_{n=1} \frac{(2\beta)^{n-1}}{n} \zeta_{\rm G}({\bf x})^n\,, \qquad \left( \mbox{ when } |\zeta_{\rm G}({\bf x})| < \frac{1}{2\beta} \,\,\right)\,.
\eea
This non-perturbative relation can be motivated by single-field models with a non-stationary inflection point that triggers a USR phase~\cite{Atal:2018neu,Tomberg:2023kli}, in curvaton models where the curvaton dominates the energy budget during its decay ($r_{\rm dec} \to 1$)~\cite{Sasaki:2006kq}, and in models of slow first-order phase transitions~\cite{Lewicki:2024ghw}. To enable a viable non-perturbative treatment, we must extend the description of the non-Gaussian field into the region $|2\beta \zeta_{\rm G}({\bf x})| \geq 1$. Following Ref.~\cite{Iovino:2024sgs}, we introduced the absolute value under the logarithm. This approach approximates well the NG of the curvaton~\cite{Sasaki:2006kq} (in the $r_{\rm dec} \to 1$ limit; see also~\cite{Pi:2021dft,Pi:2022ysn}). However, for USR models, lattice simulations indicate a different continuation beyond $2\beta\zeta_{G} \gtrsim 1$~\cite{Caravano:2025diq}. As such, the example~\eqref{eq:F_USR_abs} is meant to serve as an illustration of the formalism above rather than a complete study of NG in models listed above, for which accounting for deviations from Eq.~\eqref{eq:F_USR_abs} might be necessary.

The coefficients of the local expansion corresponding to \eqref{eq:F_USR_abs} are 
\be
    F_{{\rm NL},n} = \frac{1}{n}(2\beta)^{n-1}.
\ee
Already at this stage, it is possible to predict potential problems regarding perturbation theory: the logarithmic series converges notoriously slowly, and the series expansion breaks down when the size of the fluctuations exceeds $1/(2\beta)$, which is always a possibility with Gaussian fluctuations. In general, since $F$ is non-analytic, the local expansion will inevitably fail for sufficiently strong fluctuations. Nevertheless, a (semi-)perturbative approach still exists once we consider the resummed expansion \eqref{eq:NP_Cn_expansion}. For notational brevity, we can absorb $\xi_0$ into $\beta$ via the following  redefinitions
\be\label{eq:USR_rescalings}
    \bar\beta = \sqrt{\xi_0} \beta\,,\quad
    \psi_{ij} \equiv \xi_{ij}/\xi_0\,, \quad
    \bar\zeta_{\rm G} \equiv \zeta_{\rm G}/\sqrt{\xi_0}\,, \quad
    \bar\zeta \equiv \zeta/\sqrt{\xi_0}\,.
\ee
Now, the amount of NG is quantified by the parameter $\bar\beta$, the expansion \eqref{eq:NP_Cn_expansion} takes the form
\be
    \left\langle\prod_i \bar\zeta({\bf x}_i)\right\rangle
    = \sum_{\nu_{ij} \in \mathcal{N}}  \frac{\prod_{i<j} \psi_{ij}^{\nu_{ij}}}{\prod_{i<j} \nu_{ij}!} \prod_i s_i!\bar{\mathcal{C}}_{s_i}\,.
\ee
and the rescaled coefficients $\bar{\mathcal{C}}_{s}$ are given by
\bea\label{eq:Cn_USR}
    \bar{\mathcal{C}}_{s} 
    &\equiv \mathcal{C}_{s} \xi_0^{(s-1)/2} 
    = - \frac{1}{2\bar\beta}\frac{1}{s!}\left\langle \tilde{H}_{s}(\bar\zeta_{\rm G}({\bf x})) \ln|1-2\bar\beta \bar\zeta_{\rm G}({\bf x})|\right\rangle\,,
\eea
By Eq.~\eqref{eq:Cn_expansion} the $\bar{\mathcal{C}}_{s}$ coefficients could further be expanded in $\bar\beta$ giving
\be\label{eq:Cn_USR_pert}
    \bar{\mathcal{C}}_{s}  = \sum_{k\geq 0} \frac{(s+2k-1)!}{s!(2k)!!}  (2\bar\beta)^{s+2k-1} \,
\ee
The radius of convergence of this series is 0, so it can only be interpreted as an asymptotic series. 

\begin{figure}[t]
\centering
\includegraphics[width=0.45\textwidth]{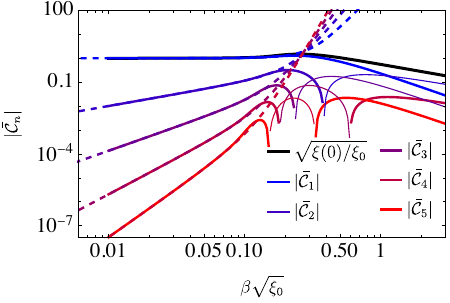}
\includegraphics[width=0.45\textwidth]{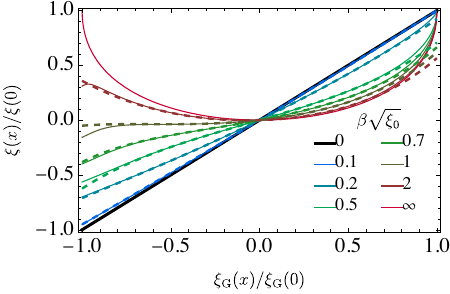}
\caption{\emph{Left panel:} The first 5 coefficients $\bar{\mathcal{C}}_n$ for USR-like models estimated as the average \eqref{eq:Cn_USR} (solid) numerically and from the first 4 terms in the perturbative expansion \eqref{eq:Cn_USR_pert} (dashed). The solid line depicts the dispersion $\sqrt{\xi(0)/\xi_0}$ of the non-Gaussian field. \emph{Right panel:} The map $G_2$ for USR-like models for different $\bar \beta$ shown in the figure. The solid lines show the exact numerical estimates using Eq.~\eqref{eq:NG_2P_function} and the dashed lines the expansion~\eqref{eq:Cn_USR_pert} in $\mathcal{C}_n$ containing the first 5 terms.} 
\label{fig:maps}
\end{figure}

The coefficients $\bar{\mathcal{C}}_{s}$ are shown in the left panel Fig.~\ref{fig:maps}. The exact numerical (solid) and perturbative estimates (dashed) using the first 4 terms in the expansion show good agreement when
\be
    \bar\beta \equiv \beta \xi_0 < \mathcal{O}(10^{-1})
\ee
and diverge rapidly for larger values. This implies that perturbative estimates can fail much before the parameter $\bar\beta$, which controls non-perturbative effects, reaches unity. To contrast this with more conventional notation, $\xi_0$ can be roughly compared to the height of the dimensionless power spectrum $\xi_0 \approx \mathcal{P}_{\zeta}(k_{\rm max})$ and $f_{\rm NL} \equiv (3/5)F_{\rm NL,2} = (3/5)\beta$. we find that perturbativity fails when $\mathcal{P}_{\zeta}(k_{\rm max}) f_{\rm NL}^2 < \mathcal{O}(10^{-2})$. Relying on the validity of asymptotic series, we found a slightly higher perturbativity bound $\xi_0 f_{\rm NL}^2 < 0.08$ in Ref.~\cite{Iovino:2025cdy}. A qualitatively similar conclusion was reached in Ref.~\cite{Iovino:2024sgs} using non-perturbative arguments that differ from the ones presented here.

The mapping $\mathcal{G}_2$ of the two-point function is shown in the right panel of Fig.~\ref{fig:maps} for different $\bar\beta$. As explained above, the mapping $\mathcal{G}_2$ is independent of the shape of the power spectrum. For USR-like models \eqref{eq:F_USR_abs}, the explicit dependence of the amplitude also drops out in the ratios $\xi(r)/\xi(0)$ and $\xi_{\rm G}(r)/\xi_{\rm G}(0) \equiv \psi(r)$, so the result depends only on a single parameter $\bar \beta$. The solid lines show the exact estimate~\eqref{eq:NG_2P_function}~\footnote{Numerically, we evaluate
\bea
    \mathcal{G}_2(\psi)
    =& \frac{1}{4\pi \bar\beta^2 \sqrt{1-\psi^2}}\int^{\infty}_{-\infty} \td \zeta_{+} \td \zeta_{-} \ln|1-2\bar\beta (\zeta_{+}\!+\!\zeta_{-})| 
    \ln|1-2\bar\beta (\zeta_{+}\!-\!\zeta_{-})| e^{- \frac{\zeta_{+}^2}{1+\psi} - \frac{\zeta_{-}^2}{1-\psi}}
    \\
    &- \frac{1}{8 \pi \bar \beta^2 }\left[\int^{\infty}_{-\infty} \td \bar\zeta_{\rm G} \ln|1-2\bar\beta \bar\zeta_{\rm G}| e^{- \frac{\zeta_{\rm G}^2}{2}}\right]^2\,,
\eea
where we redefined the variables of integration $\zeta_{\pm} \equiv (\bar \zeta_1 \pm \bar \zeta_2)/2$ and subtracted the mean $\langle F(\zeta_{\rm G})\rangle^2$, which does not generally vanish for the ansatz~\eqref{eq:F_USR_abs}. This is equivalent to performing a shift in the NG field that guarantees~\eqref{eq:vanishing_mean}. The NG two-point function is then $\mathcal{G}_2(\psi(r))$ and the power spectrum is obtained by numerically computing the Fourier transform~\eqref{eq:xi_to_P} of $\mathcal{G}_2(\psi(r))$.
}
while the dashed lines show the expansion~\eqref{eq:2P_expansion_xi12} in $\mathcal{C}_s$ up to order 5 (i.e., $\xi_{G}(r)^5$). Unlike the expansion in $F_{{\rm NL},n}$ \eqref{eq:Cn_USR_pert}, the expansion in $\mathcal{C}_s$~\eqref{eq:2P_expansion_xi12} can capture the non-linearities when $\bar\beta \gtrsim 1$. The largest deviations can be observed when $|\xi_{\rm G}(r)/\xi_0| \approx 1$, that is, when the Gaussian field is strongly (anti)correlated, which is expected, as the \eqref{eq:NP_Cn_expansion} is a power law expansion in $\xi_{\rm G}(r)$.

\paragraph{Large $\bar\beta$ limit.} Although the series \eqref{eq:Cn_expansion} in $F_{{\rm NL},n}$ giving the coefficients $\mathcal{C}_s$ diverges, we can evaluate these coefficients directly from Eq.~\eqref{eq:Cn_USR} and obtain the convergent series \eqref{eq:2P_expansion_xi12} for $\mathcal{G}_2$. This can be demonstrated explicitly in the strong NG limit $\bar\beta \to \infty$, where $\mathcal{G}_2$ and thus also $\mathcal{C}_s$ can be found analytically. \footnote{In the $\bar\beta = \mathcal{O}(1)$ regime, we checked the convergence of \eqref{eq:2P_expansion_xi12} numerically.} In that limit, we can approximate
\be
    \zeta({\bf x}) 
    = - \frac{1}{2\bar\beta}\ln\left|2\bar\beta \bar\zeta_{\rm G}({\bf x})\right| + \mathcal{O}(\bar\beta^{-2})\,,
\ee
where the $2\bar\beta$ can be absorbed by a shift, so $\bar\beta$ will only appear in the prefactor. The two point function can be obtained from the correlation of the logarithms of the Gaussian field $\langle \ln(\bar\zeta_1)\ln(\bar\zeta_2 \rangle - \langle \ln(\bar\zeta)\rangle^2$, which gives\footnote{The NG correlation function is given by Eq.~\eqref{eq:NG_2P_function}, which can be expressed by the double integral
\bea
    \frac{\mathcal{G}_2(\psi_{12})}{\xi_0}  
    &=  \frac{1}{(2\bar \beta)^2}\int_{\zeta_i \in \mathbb{R} } \frac{\td \bar\zeta_1 \td \bar\zeta_2}{2\pi \sqrt{1 - \psi_{12}^2}} \log(|\zeta_1|) \log(|\zeta_2|) \left[ e^{- \frac{1}{2} \frac{(\bar\zeta_1^2 + \bar\zeta_2^2) - 2\bar\zeta_1 \bar\zeta_2 \psi_{12} }{1 - \psi_{12}^2} } 
    \!-\! e^{- \frac{1}{2} (\bar\zeta_1^2 + \bar\zeta_2^2)} \right] \,
    \\
    &=  \frac{4}{(2\bar \beta)^2} \partial_{\alpha_1} \partial_{\alpha_2} \left. \int_{\zeta_i \geq 0} \frac{\td \bar\zeta_1 \td \bar\zeta_2}{2\pi \sqrt{1\!-\!\psi_{12}^2}} \bar\zeta_1^{\alpha_2} \bar\zeta_2^{\alpha_2}
    \left[
    \cosh\left(\frac{2\psi_{12}}{1\!-\!\psi_{12}^2} \bar\zeta_1 \bar\zeta_2 \right)
     e^{- \frac{1}{2} \frac{(\bar\zeta_1^2 + \bar\zeta_2^2)}{1\!-\!\psi_{12}^2}}
     \!-\! e^{- \frac{1}{2} (\bar\zeta_1^2 + \bar\zeta_2^2)} \right]\right|_{\alpha_i = 0}\,
     \\
     &= \frac{1}{(2\bar \beta)^2} \partial_{\alpha_1} \partial_{\alpha_2} \left.{}_2F_1(-\alpha_1/2,-\alpha_2/2,1/2,\psi_{12}^2)\right|_{\alpha_i = 0}\,,
     \nonumber
\eea
where the last two lines give a rough outline how the integral can be evaluated by first replacing $\log(|\zeta_i|)$ by $\zeta_i^{\alpha_i}$ to obtain an expression in terms of the hypergeometric function $_2 F_1$ from which \eqref{eq:xi_USR_large_beta} can be derived. }
\be\label{eq:xi_USR_large_beta}
    \xi(x)
    \equiv \mathcal{G}_2(\xi_{\rm G}(x))
    \stackrel{\bar\beta \gg 1}{=} \frac{\xi_0}{8\bar \beta^2} \arcsin^{2}\left(\frac{\xi_{\rm G}(x)}{\xi_0}\right)\,.
\ee
Using \eqref{eq:2P_expansion_xi12}, the coefficients $\bar{\mathcal{C}}_n$ can be obtained by expanding $\mathcal{G}_2$ in $\xi_{\rm G}(x)$ up to their sign, which must be resolved from the definition~\eqref{eq:Cs_integral}. They are
\be\label{eq:USR_Cn_asymp}
    \bar{\mathcal{C}}_n
    \stackrel{\bar\beta \gg 1}{=} \frac{1}{2\bar\beta}
    \left\{\begin{array}{cc}
    -\frac{\sqrt{\pi} \, (-2)^{-n/2}}{n \Gamma((n+1)/2)}    & \quad n \mbox{ even}\\
    0    & \quad n \mbox{ odd}
    \end{array}\right.\,.
\ee
Since all coefficients are suppressed by the same factor $1/\bar\beta$ in this regime, the parameter $\bar\beta$ ceases to be a good quantifier of NG. As expected, the normalized correlation function
\be
    \frac{\xi(x)}{\xi(0)}
    \stackrel{\bar\beta \gg 1}{=} \frac{4}{\pi^2} \arcsin^{2}\left(\frac{\xi_{\rm G}(x)}{\xi_0}\right)\,
\ee
(and the power spectrum) is independent of $\bar\beta$ and the effect of NG can be considered to be saturated. This case is shown by the red curve in Fig.~\ref{fig:maps}.

\paragraph{Power spectrum.} 

\begin{figure}[t]
\centering
\includegraphics[width=\textwidth]{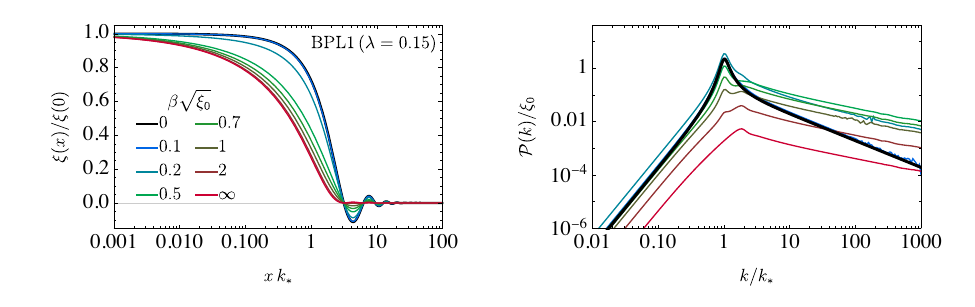}
\vspace{-6mm}
\caption{\emph{Left panel:} The deformation of the 2-point function $\mathcal{P}_{\rm G,1}(k)$ in Eq.~\eqref{eq:PG_templates} (black line) due to varying amount of local NG (colored lines). \emph{Right panel:} The deformation of the Gaussian power spectrum (black line) due to varying amount of local NG (colored lines). The normalization of the $\bar\beta \to \infty$ power spectrum is arbitrary. } 
\label{fig:BPL1}
\end{figure}

\begin{figure}[t]
\centering
\includegraphics[width=\textwidth]{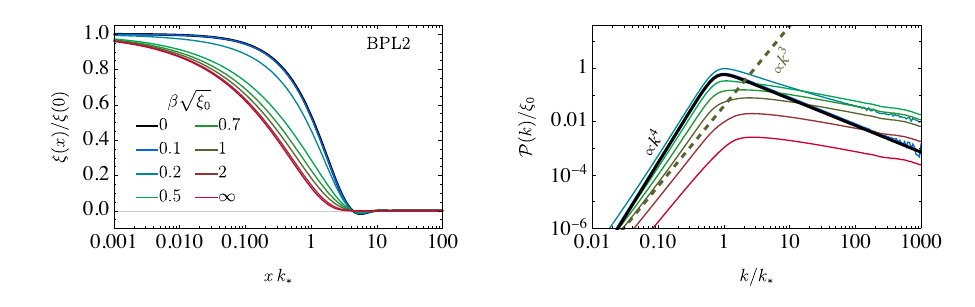}
\vspace{-6mm}
\caption{Same as Fig.~\ref{fig:BPL1}, but for the template $\mathcal{P}_{\rm G,2}(k)$ in Eq.~\eqref{eq:PG_templates}. The dashed line indicates $k^3$ growth.} 
\label{fig:BPL2}
\end{figure}

The effect of the NG on the two-point function and the power spectrum for different $\bar\beta$ is illustrated in Figs.~\ref{fig:BPL1} and Fig.~\ref{fig:BPL2}. In these figures, we assumed the following Gaussian power spectra:
\bea\label{eq:PG_templates}
    \mathcal{P}_{\rm G,PBL1}(k)/\xi_0 
    \!=\!\frac{4\mu}{\pi}\frac{k^3}{(k^2\!-\!k_*^2 + \mu^2)^2\!+\!4 k_*^2 \mu^2}\,,
    \quad
    \mathcal{P}_{\rm G,PBL2}(k)/\xi_0 
    \!=\!\frac{2.86}{(k/k_*)^{-4}\!+\!4 (k/k_*)}\,,
\eea
with $\mu = 0.15$. Both spectra have $k^{-1}$ UV tails and $k^3$ and $k^4$ IR tails, depicted by the black curves in Fig.~\ref{fig:BPL1} and Fig.~\ref{fig:BPL2}. With these templates, the correlation functions $\xi_{\rm G}(r)$ can be obtained analytically, which helps to reduce numerical noise. 

In both situations, the power spectrum is first amplified by the NG, but as the NG is increased further, the power spectrum becomes suppressed. Perturbatively, the NG power spectrum is given by a sum of positive terms \eqref{eq:2P_expansion_PG}, which means that the coefficients $\mathcal{C}_s$ must decrease as $\bar \beta$ grows. This effect can be observed in Fig.~\eqref{fig:maps} and agrees well with the asymptotic~\eqref{eq:USR_Cn_asymp}.

The normalisation of the power spectrum is encoded in $\xi(0)/\xi_0 = \sum_{n} n! \bar{\mathcal{C}}_n^2$, shown by the black line in the left panel of Fig.~\eqref{fig:maps}. The $\bar \beta \gg 1$ limit gives
$
    \xi(0)/\xi_0
    \sim \pi^2/(32\bar\beta^2).
$
by Eq.~\eqref{eq:xi_USR_large_beta} and it agrees well with numerical estimates already when $\bar\beta \gtrsim 0.5$.


We further observe the flattening of the peak and a shallower spectral slope at $k \gg k_*$. This is due to a manifestly non-perturbative effect, which we will discuss in detail below. Some features of the peak seem to be retained for the BPL1 case shown in Fig.~\ref{fig:BPL1}, where the Gaussian spectrum possesses a resonant peak. Fig.~\ref{fig:BPL2} demonstrates a transition from a $k^4$ to a $k^3$ IR-tail as $\bar\beta$. The formation of $k^3$ tails for non-Gaussian fields was expected based on the reasoning surrounding Eq.~\eqref{eq:k3_tail}.

Figs.~\ref{fig:BPL1},~\ref{fig:BPL2} illustrate how the power spectrum of the non-Gaussian field $\zeta$ changes when the non-Gaussian field is constructed from a Gaussian field $\zeta_{\rm G}$ with a \emph{fixed} power spectrum. However, one can also ask whether it is possible to recover the Gaussian power spectrum, given that we know the power spectrum of the non-Gaussian field. As the relation between correlation functions is fully covered by the map $\mathcal{G}_2$ shown in~\eqref{fig:maps} which, by our numerical estimates, is invertible when $\bar\beta \lesssim 1$ for USR-like models. Thus, it is in principle possible to construct the Gaussian correlation function $\xi_{\rm G}(x) = \mathcal{G}^{-1}_2(\xi(x))$ and recover the power spectrum in that way. However, as argued in \eqref{sec:loc_NG_basics}, this estimate may not be consistent, as the relation \eqref{eq:F_USR_abs} between the Gaussian and non-Gaussian fields is not invertible, and thus the resulting $n$-point functions could display NG features.

\paragraph{Flattening of the UV-tail.}

To gain further insight into the UV-tail, it is necessary to analyse the behaviour of $\xi(r)$ at small separations. This regime is governed by $\mathcal{G}_2$ in the strongly correlated limit, $\psi_{12} \equiv \xi_{12}/\xi_0 \to 1$. To obtain the corresponding asymptotic form, we rewrite the non-Gaussian two-point correlation function as
\be
    \langle \zeta({\bf x}_1) \zeta({\bf x}_2) \rangle 
    = \langle \zeta({\bf x}_1)^2 \rangle - Q(\psi_{12}),
\ee
where we introduce the quantity
\be
    Q(\psi_{12}) 
    \equiv \frac{1}{2}\langle (F(\zeta_1) - F(\zeta_2))^2\rangle 
    = \mathcal{G}_2(\psi_{12}=1) - \mathcal{G}_2(\psi_{12}) \geq 0\,,
\ee
which, by construction, vanishes in the $\psi_{12} \to 1$ limit. This quantity is particularly convenient in the context of the logarithmic ansatz \eqref{eq:F_USR_abs}, as it can be expressed in the form
\be
    Q(\psi_{12}) 
    = \frac{1}{8 \beta^2}\left\langle \ln^2\left| \frac{1-2\bar\beta \bar\zeta_1}{1-2\bar\beta \bar\zeta_2}\right|\right\rangle
    = \frac{1}{8 \beta^2}\left\langle \ln^2\left| \frac{z - \sqrt{1-\psi_{12}}}{z + \sqrt{1-\psi_{12}}}\right|\right\rangle ,
\ee
where we defined
\be
    z \equiv \frac{\frac{1}{\sqrt{2}\bar\beta} - \sqrt{1+\psi_{12}}\zeta_{+}}{\zeta_{-}},
    \qquad
    \zeta_{\pm} \equiv \frac{\bar\zeta_1 \pm \bar\zeta_2}{\sqrt{2 (1\pm\psi_{12})}} \, .
\ee
The random variables $\zeta_{\pm}$ are independent Gaussian variables with unit variance. When $\psi_{12} \to 1$, then the average is approximately\footnote{The relevant integrals in the derivation are
\bea
    P(z=0) 
    = \frac{1}{2 \pi \sqrt{\psi_{12}+1}} \int_{-\infty}^{\infty } \td\zeta_{-}\, | \zeta_{-}|  \exp \left(-\frac{\zeta_{-}^2}{2}-\frac{1}{2} \left(\frac{\sqrt{2}}{2 \bar\beta  \sqrt{\psi_{12}+1}}\right)^2\right) 
    &= \frac{e^{-\frac{1}{4 \bar\beta^2 (1+\psi_{12})}}}{\pi  \sqrt{1+\psi_{12}}}\,,
    \\
    \int^{\infty}_{-\infty} \td z\,  \ln^2\left| \frac{z - \sqrt{1-\psi_{12}}}{z + \sqrt{1-\psi_{12}}}\right| 
    = \sqrt{1-\psi_{12}} \int^{\infty}_{-\infty} \td z'\,  \ln^2\left| \frac{z' - 1}{z' + 1}\right| 
    &= 2\pi^2\,,
\eea
where we used the rescaling $z' = z\sqrt{1-\psi_{12}}$.
}
\bea\label{eq:Q_USR}
     Q(\psi_{12}) 
     &= \frac{1}{8 \beta^2} \int^{\infty}_{-\infty} \td z\, P(z) \ln^2\left| \frac{z - \sqrt{1-\psi_{12}}}{z + \sqrt{1-\psi_{12}}}\right|
     \\
     &\sim \frac{1}{8 \beta^2} P(z=0)\int^{\infty}_{-\infty} \td z\,  \ln^2\left| \frac{z - \sqrt{1-\psi_{12}}}{z + \sqrt{1-\psi_{12}}}\right|
     \\
     &\sim C_{\bar \beta} \sqrt{1-\psi_{12}} \,,   
     \quad \mbox{with} \quad
     C_{\bar \beta} \equiv \xi_0 \frac{\pi e^{-\frac{1}{8\bar\beta^2}}}{4 \sqrt{2} \bar \beta^2}\,,
\eea
where $P(z)$ is the distribution of $z$. The $\bar \beta$ dependence of $C_{\bar \beta}$ showcases the fundamentally non-perturbative nature of this limit.

As a result, the non-Gaussian 2-point function behaves as
\be\label{eq:psi_0_asymptotic}
    \xi(r) 
    \sim \mathcal{G}_2(\xi_0) + C_{\bar \beta} \, \sqrt{1-\xi_{\rm G}(r)/\xi_0}
\ee
near strongly correlated regions. In particular, when $r \to 0$, we can generally expand the Gaussian power spectrum as
\be
    \xi_{\rm G}(r) \sim \xi_0 - A r^{\alpha}\,,
\ee
where $A$ and $\alpha$ are positive constants related to the UV behaviour of the power spectrum. For correlation functions that is regular in $\mathbf r$, $\alpha$ must be an even positive integer. In the case of non-analytic behaviour in $\mathbf r$ is confined to the origin, the $k \to \infty$ asymptotic of the power spectrum is
\be
    \mathcal{P}_{\rm G}(k) \sim A \gamma_\alpha k^{-\alpha}\,, 
    \qquad \mbox{when} \quad
    0< \alpha \leq 2\,,
\ee
where we introduced the coefficient $\gamma_\alpha \equiv (2/\pi)\Gamma(2+\alpha)\sin\left(\pi\alpha/2\right)$. Note that this coefficient vanishes when $\alpha \in 2\mathbb Z$ is an even integer, in which case the leading small $r$ term does not tell us much about the large $k$ behaviour. 

As a result of \eqref{eq:psi_0_asymptotic}, the NG two point function behaves as
\be
    \xi(r) 
    = \mathcal{G}_2(\xi_0) + C_{\bar \beta} \, \sqrt{A} r^{\alpha/2}\,.
\ee
This implies that the tail of the non-Gaussian power spectrum is flattened
\be\label{eq:P_asymp}
    \mathcal{P}_{\zeta}(k) \sim C_{\bar \beta} \gamma_{\alpha/2} \, \sqrt{A} k^{-\alpha/2}\,.
\ee
so the spectral slope is halved. Notably, regular power spectra with 

An illustrative case is provided in Figs.~\ref{fig:BPL1} and~\ref{fig:BPL2}, where both Gaussian spectra \eqref{eq:PG_templates} exhibit a $k^{-1}$ tail, whereas the corresponding non-Gaussian spectra display a $k^{-1/2}$ tail for all curves with $\bar\beta \geq 0.2$. While this spectral flattening appears to be a robust and universal feature, it is rapidly diminished in the limit $\bar \beta \to 0$, since the normalisation of the UV-tail is proportional to
\be 
    C_{\bar \beta} \propto e^{-\frac{1}{8\bar\beta^2}}\,.
\ee
For instance, we find that $\bar{\beta} = 0.2$ yields $C_{\beta} \approx 0.6$, whereas for $\bar{\beta} = 0.1$ the coefficient decreases to $C_{\beta} \approx 2\times 10^{-4}$, rendering its signal essentially indistinguishable in Figs.~\ref{fig:BPL1},~\ref{fig:BPL2}, as its contribution is overtaken by numerical noise. For $\bar{\beta} \geq 0.2$, we have explicitly verified that the analytical approximation given in Eq.~\eqref{eq:P_asymp} is in good agreement with the numerical results shown in Figs.~\ref{fig:BPL1},~\ref{fig:BPL2}.

Finally, we remark that the $\sqrt{1-\psi_{12}}$ asymptotic obtained from \eqref{eq:Q_USR} originates from the logarithmic singularity. In physically realistic situations, however, it is natural to require $\zeta$ to remain finite. Implementing a regularisation of the USR-like ansatz \eqref{eq:F_USR_abs} to eliminate these divergences will necessarily modify the behaviour in the limit $r \to 0$. Nonetheless, provided that $\zeta$ is allowed to increase logarithmically, approaching $\zeta_{\rm G} \to 1/(2\beta)$ by a sufficiently large margin, one expects the existence of an intermediate regime in which the scaling $\mathcal{P}_{\zeta} \propto k^{-\alpha/2}$ is retained.

For instance, obtaining the numerical scaling $\mathcal{P}_{\zeta} \propto k^{-1/2}$ in Figs.~\ref{fig:BPL1} and~\ref{fig:BPL2} requires very high numerical resolution of $\mathcal{G}_2$ in the vicinity of $\psi_{12} = 1$. This is because computing $\mathcal{G}_2$ on a grid of $\psi_{12}$ and subsequently applying linear interpolation leads to $Q \propto 1 - \psi_{12}$ as $\psi_{12} \to 1$, and consequently induces an artificial $\mathcal{P}_{\zeta} \propto k^{-1}$ tail at sufficiently large $k$. To eliminate such numerical artefacts and correctly capture the $k^{-1/2}$ asymptotic behaviour, we must employ a sufficiently fine grid in $\psi_{12}$ for the calculations underlying Figs.~\ref{fig:BPL1} and~\ref{fig:BPL2}, or alternatively enforce Eq.~\eqref{eq:Q_USR} explicitly in the limit $\psi_{12} \to 1$. In this case, the discretization of $\psi_{12}$ effectively acts as a regulator, modifying the ultraviolet tail at the largest values of $k$ while preserving the $k^{-1/2}$ scaling over a broad intermediate range of $k$. It is reasonable to expect that physically motivated regularisation mechanisms exert a qualitatively similar influence on the UV tail, although a detailed investigation of such effects lies beyond the scope of the present study.


\section{Summary}
\label{sec:concl}

We studied the statistical properties of statistically homogeneous and isotropic locally non-Gaussian fields, that is, random fields $\zeta({\bf x})$ that can be constructed from Gaussian fields $\zeta_{\rm G}({\bf x})$ via a local transformation $\zeta({\bf x}) = F(\zeta_{\rm G}({\bf x}))$ that depends only on the field value at a given point.

We demonstrated that the $n$-point functions of such fields can be completely characterised by \emph{i)} a (possibly infinite) set of coefficients $\mathcal{C}_s$, which depend only on the one-point distribution of the non-Gaussian field, and  \emph{ii)} the correlation function of the auxiliary Gaussian field. To exploit this feature, we developed a perturbative diagrammatic framework that does not rely on the commonly used series expansion of $F(\zeta_{\rm G})$ and can also be applied when $F(\zeta_{\rm G})$ is non-analytic. This is possible because the multivariate Gaussian distribution can be decomposed using the Kibble–Slepian formula, which allows us to reduce all non-Gaussian contributions from $F(\zeta_{\rm G})$ to one-dimensional Gaussian averages. When $F$ is analytic, it can be shown that this procedure can be understood as a resummation of "tadpole" diagrams.

We showed that the $n$-point functions of the non-Gaussian field can be constructed via functions $\mathcal{G}_n$ that map Gaussian 2-point functions to non-Gaussian $n$-point functions. The $\mathcal{G}_n$ functions can be defined independently of the shape of the Gaussian power spectrum, and they only depend on its amplitude and $F(\zeta_{\rm G})$. In practice, this allows us to speed up the computation of non-Gaussian $n$-point functions in the non-perturbative regime for a broad class of Gaussian power spectra, since the map $\mathcal{G}_n$ needs to be evaluated only once for a given $F(\zeta_{\rm G})$. However, more progress is needed before fully non-perturbative $n \geq 3$ point functions can be obtained in a fast and reliable manner, since the parameter space of covariance matrices can grow prohibitively large quite rapidly when $n \geq 3$. Such progress would be needed, for instance, to estimate SIGW spectra in a strongly non-Gaussian regime.

As a case study, we examined local non-Gaussianity that generates exponential tails in the distribution of one-point fluctuations. We derived the mapping between the correlation functions and power spectra of the Gaussian and non-Gaussian fields and determined the non-perturbative coefficients relevant for the resummed perturbative framework. While these maps and the resummed coefficients had to be evaluated numerically in the general case, we were able to obtain exact analytic expressions in the limit of strong non-Gaussianity. We also observed that this type of non-Gaussianity tends to flatten non-Gaussian power spectra, and once the non-Gaussianity becomes sufficiently large, it can cause a substantial reduction in the amplitude of the power spectrum and give rise to $\mathcal{P}_{\rm zeta} \propto k^3$ IR tails.

\acknowledgments
\noindent
The author thanks G. Perna for productive discussions and for feedback on the manuscript, and G. Franciolini, A. Iovino, and C. \"Unal for useful discussions. This work is supported by the Estonian Research Council grants PSG869, TARISTU24-TK3, TARISTU24-TK10, and the Center of Excellence program TK202.

\bibliographystyle{JHEP}
\bibliography{refs}

\end{document}